\documentclass[aps,twocolumn,superscriptaddress]{revtex4}
\usepackage{graphicx}
\begin{document}
\bibliographystyle{apsrev}


\title{Slater Transition in the Pyrochlore Cd$_{2}$Os$_{2}$O$_{7}$}



\author{D. Mandrus}
\email{mandrusdg@ornl.gov}
\affiliation{Solid State Division, Oak Ridge National Laboratory, Oak Ridge, TN 37831}
\affiliation{Department of Physics, The University of Tennessee, Knoxville, TN 37996}
\author{J. R. Thompson}
\affiliation{Department of Physics, The University of Tennessee, Knoxville, TN 37996}
\affiliation{Solid State Division, Oak Ridge National Laboratory, Oak Ridge, TN 37831}

\author{R. Gaal}
\affiliation{\'{E}cole Polytechnique Federale de Lausanne, Department de Physique, CH-1015, Lausanne,
Switzerland}

\author{L. Forro}
\affiliation{\'{E}cole Polytechnique Federale de Lausanne, Department de Physique, CH-1015, Lausanne,
Switzerland}

\author{J. C. Bryan}
\affiliation{Chemical and Analytical Sciences Division, Oak Ridge National Laboratory, Oak Ridge, TN
37831}

\author{B. C. Chakoumakos}
\affiliation{Solid State Division, Oak Ridge National Laboratory, Oak Ridge, TN 37831}

\author{L. M. Woods}
\affiliation{Department of Physics, The University of Tennessee, Knoxville, TN 37996}
\affiliation{Solid State Division, Oak Ridge National Laboratory, Oak Ridge, TN 37831}

\author{B. C. Sales}
\affiliation{Solid State Division, Oak Ridge National Laboratory, Oak Ridge, TN 37831}

\author{R. S. Fishman}
\affiliation{Solid State Division, Oak Ridge National Laboratory, Oak Ridge, TN 37831}

\author{V. Keppens}
\altaffiliation [Permanent address: ]{National Center for Physical Acoustics and Department of
Physics, The University of Mississippi, Oxford, MS 38677} \affiliation{Solid State Division, Oak
Ridge National Laboratory, Oak Ridge, TN 37831}

\date{\today}

\begin{abstract}
$\mathrm{Cd_2Os_2O_7}$ crystallizes in the pyrochlore structure and undergoes a metal-insulator
transition (MIT) near 226 K.  We have characterized the MIT in $\mathrm{Cd_2Os_2O_7}$ using X-ray
diffraction,  resistivity at ambient and high pressure, specific heat, magnetization, thermopower,
Hall coefficient, and thermal conductivity.  Both single crystals and polycrystalline material were
examined.  The MIT is accompanied by no change in crystal symmetry and a change in unit cell volume
of less than 0.05\%. The resistivity shows little temperature dependence above 226 K, but increases
by 3 orders of magnitude as the sample is cooled to 4 K. The specific heat anomaly resembles a
mean-field transition and shows no hysteresis or latent heat. $\mathrm{Cd_2Os_2O_7}$ orders
magnetically at the MIT.  The magnetization data is consistent with antiferromagnetic order, with a
small parasitic ferromagnetic component.  The Hall and Seebeck coefficients are consistent with a
semiconducting gap opening at the Fermi energy at the MIT.  We have also performed electronic
structure calculations on $\mathrm{Cd_2Os_2O_7}$.  These calculations indicate that
$\mathrm{Cd_2Os_2O_7}$ is metallic, with a sharp peak in the density of states at the Fermi energy.
We intepret the data in terms of a Slater transition.  In this scenario, the MIT is produced by a
doubling of the unit cell due to the establishment of antiferromagnetic order.  A Slater
transition---unlike a Mott transition---is predicted to be continuous, with a semiconducting energy
gap opening much like a BCS gap as the material is cooled below $T_{MIT}$.
\end{abstract}

\maketitle

\section{Introduction}

Oxides containing second and third row ($4d/5d$) transition metals such as Mo, Ru, Re, Os, and Ir
display an impressive variety of phenomena and are a good place to search for interesting new
materials. Electrically, these $4d/5d$ materials range from excellent metals such as ReO$_{3}$ with a
room temperature conductivity about the same as Cu \cite{Sleight66}, to Mott-Hubbard insulators such
as $\mathrm{Ca_2RuO_4}$ \cite{Maeno99} and $\mathrm{Y_2Ru_2O_7}$ \cite{Cox83}. Magnetically, we find
an equally wide range of behavior, from local-moment antiferromagnetism ($\mathrm{Sr_2YRuO_6}$
\cite{Cox83}) to weak ferromagnetism ($\mathrm{Sr_2IrO_4}$ \cite{Cao98}) to spin-glass behavior
($\mathrm{Co_2RuO_4}$ \cite{Mandrus99}) to itinerant ferromagnetism ($\mathrm{SrRuO_3}$
\cite{Longo71}). Some novel collective phenomena have also been discovered in $4d/5d$ materials, such
as the definitely unusual (and perhaps $p$-wave) superconductivity in $\mathrm{Sr_2RuO_4}$
\cite{Maeno94}, and the coexistence of antiferromagnetism and superconductivity in
$\mathrm{RuSr_2GdCu_2O_8}$ \cite{Bernhard99} \cite{Lynn2000}.

The unique properties of ${4d/5d}$ materials stem from the properties of the $4d/5d$ electrons
themselves, which are much less localized than $3d$ electrons and play a greater role in chemical
bonding. Because of the increased hybridization with the coordinating O$^{2-}$ anions, $4d/5d$ ions
are typically found in low-spin (large crystal field) configurations rather than the high-spin
configurations typical of $3d$ ions. The larger spatial extent of $4d/5d$ wavefunctions also means
that the intra-atomic Coulomb repulsion is smaller and therefore $4d/5d$ materials are generally less
strongly correlated than $3d$ materials. In theoretical treatments of electron correlation, a measure
of the degree of correlation is given by the ratio $U/W$, where $U$ is the intra-atomic Coulomb
repulsion (``Hubbard $U$'') and $W$ is the bandwidth. Typically, for $3d$ oxides $U/W \gg 1$ and
these materials are thought to be in the ``strong coupling limit.'' For simple metals, on the other
hand, $U/W \ll 1$ and the ``weak coupling limit'' applies. $4d/5d$ oxides represent an intermediate
case, however, with $U/W \approx 1$. This ``intermediate coupling regime'' has been much less
studied, both experimentally and theoretically, and the exploration of this parameter space along
with the expectation of uncovering new phenomena is the primary motivation for studying this class of
materials.

The synthesis and initial characterization of $\mathrm{Cd_2Os_2O_7}$ were reported in 1974 by
Sleight, \textit{et al.} \cite{Sleight74}, but no subsequent publications have appeared on this
compound. The physical properties reported in Ref. \onlinecite{Sleight74} are quite intriguiging:
$\mathrm{Cd_2Os_2O_7}$ was found to crystallize in the pyrochlore structure and to undergo a
continuous, purely electronic metal-insulator transition (MIT) near 225 K. It was also found that the
MIT was coincident with a magnetic transition that the authors characterized as antiferromagnetic.

In this work, we present a comprehensive re-investigation of the basic physical properties of
$\mathrm{Cd_2Os_2O_7}$, and have added considerable detail to the basic picture presented in Ref.
\onlinecite{Sleight74}. We find that the transition is, indeed, continuous, and that the lattice
appears to be playing no discernable role in the transition. The magnetism is clearly more complex
than a simple N\'{e}el state, and almost certainly involves the formation of a ``strong'' spin
density wave (SDW). The idea that antiferromagnetic ordering can double the unit cell and for a
half-filled band produce a metal-insulator transition goes back to Slater, who in 1951 proposed a
split-band model of antiferromagnetism \cite{Slater51}. In this model, the exchange field favors up
spins on one sublattice and down spins on another sublattice and for large $U/W$ reduces to the
atomic model of an antiferromagnetic insulator, with a local moment on each site. The thermodynamics
of the metal-insulator transition predicted by Slater's model were worked out in a mean field
approximation by Matsubara and Yokota (1954) \cite{Matsubara54}, and by Des Cloizeaux (1959)
\cite{DesCloizeaux59}. In both treatments a continuous metal-insulator transition was predicted, with
a semiconducting gap opening much the way a BCS gap opens in a superconductor.

These results have, of course, been largely supplanted by modern SDW theory \cite{Fazekas99}, but it
must be remembered that SDW theory is unambiguously effective only in the weak coupling limit. For
example, at half filling it should be possible to proceed smoothly from a weak coupling SDW state to
a Mott insulating state as $U/W$ is increased. Although strong coupling SDW theory gets some of the
aspects of the Mott insulating state right, it is wrong in other respects and must be regarded as an
incomplete description \cite{Fazekas99}.

One of the major failings of strong coupling SDW theory is that the gap is predicted to disappear
above the N\'{e}el temperature, whereas in true Mott insulators like CoO the gap persists despite the
loss of long range magnetic order. In such materials a Mott-Hubbard description is clearly correct,
but when the metal-insulator transition temperature and the  N\'{e}el temperature coincide one should
give careful thought as to whether a Mott-Hubbard or a Slater description is most appropriate. In
such cases the thermodynamics of the MIT can provide valuable information about the underlying
mechanism, because a Mott transition is expected to be discontinuous whereas a Slater transition
should be continuous \cite{Mott90}. Experimentally, all of the temperature-driven MITs that have been
studied in detail (e.g., $\mathrm{VO_2}$, $\mathrm{V_2O_3}$, $\mathrm{Fe_3O_4}$, NiS,
NiS$_{2-x}$Se$_{x}$, PrNiO$_{3}$) have been found to be discontinuous \cite{Imada98}, and this
explains---at least in part---why the effects of magnetic ordering on the MIT have been largely
discounted in these materials compared to the effects of the Coulomb interaction. In
$\mathrm{Cd_2Os_2O_7}$, on the other hand, the MIT is continuous and coincides with the  N\'{e}el
temperature. The Slater mechanism, then, is almost certainly at work in $\mathrm{Cd_2Os_2O_7}$.

\section{Crystal Chemistry}

$\mathrm{Cd_2Os_2O_7}$ belongs to a family of cubic materials known as pyrochlores
\cite{Subramanian83}. Pyrochlores contain 2 nonequivalent anion positions, and so oxide pyrochlores
have the general formula $\mathrm{A_2B_2O_6O'}$. Pyrochlores contain 8 formula units (88 atoms) per
unit cell, and belong to the space group $Fd\overline{3}m$ . There are two types of coordination
polyhedra in the pyrochlore structure, and these are loosely referred to as ``octahedral'' and
``cubic'' although this is not exactly correct because the conditions for the existence of perfect
octahedral and perfect cubic coordination cannot be simultaneously satisfied. In
$\mathrm{Cd_2Os_2O_7}$, our crystallographic results (section IV) indicate that the Os$^{5+}$ ions
have nearly perfect octahedral coordination, whereas the Cd$^{2+}$ ions are located within
scalenohedra (distorted cubes).

Several compounds with the formula $\mathrm{Cd_2M_2O_7}$ are known where M$^{5+}$ = Nb, Ta, Re, Ru,
and Os \cite{Subramanian83, Donohue65, Wang98}.  Insulating behavior is found for Nb$^{5+}$ ($4d^0$)
and Ta$^{5+}$ ($5d^0$) as would be expected for a compound with no $d$ electrons in its $t_{2g}$
manifold. Metallic behavior is observed, however, when M$^{5+}$  is Re$^{5+}$ ($5d^2$) and Ru$^{5+}$
($4d^3$). The physical properties of nearly all of these materials are interesting.
$\mathrm{Cd_2Nb_2O_7}$ is a a complex ferroelectric that undergoes a series of closely spaced
transitions \cite{Subramanian83}. $\mathrm{Cd_2Ta_2O_7}$ undergoes a structural phase transition at
204 K that has yet to be fully characterized \cite{Sleight76}. $\mathrm{Cd_2Ru_2O_7}$ is a high
pressure phase that displays as yet unexplained anomalies in its resistivity and thermal expansion
coefficient \cite{Wang98}.

From the discussion above it is clear that the $t_{2g}$ manifold of $d$ electrons is intimately
involved in the electronic conduction processes in $\mathrm{Cd_2M_2O_7}$ compounds as would be
expected from an elementary ionic picture of these materials.  Because Os$^{5+}$ is in the $5d^3$
configuration and the $t_{2g}$ manifold holds 6 electrons, it is reasonable to assume that the
conduction band in $\mathrm{Cd_2Os_2O_7}$ is close to half filling.  This condition, of course, is
required for a Slater picture of the MIT to be valid.

\section{Sample Preparation}

Both single crystals and polycrystalline material were prepared.  All preparations were performed
under a fume hood because of the extreme toxicity of OsO$_4$.  This oxide of osmium is extremely
volatile and melts at about 40 $^{\circ}$C.  Exposure of the eyes to OsO$_4$ should be strictly
avoided as permanent blindness can result.  Even brief exposure to the vapor is dangerous.  According
to the MSDS: ``If eyes are exposed to the vapor over a short period of time, night vision will be
affected for about one evening.  One will notice colored halos around lights.''

Single crystals of $\mathrm{Cd_2Os_2O_7}$ were grown following a method similar to Ref.
\onlinecite{Sleight74} by sealing appropriate quantities of CdO, Os, and KClO$_3$ in a silica tube
and heating at 800 $^{\circ}$C for 1 week. Shiny black octahedral crystals, up to 0.7 mm on an edge,
grew on the walls of the tube.  These crystals presumably grew from the vapor with OsO$_4$ as a
transport agent. Crystals from several growths were used in the experiments reported here.

Polycrystalline material was synthesized from CdO and OsO$_2$ powders.  These materials were ground
together thoroughly, pressed into pellets, and sealed in a silica tubes to which an appropriate
amount of KClO$_{3}$ was added to provide the required oxygen.  The tubes were then heated at 800
$^{\circ}$C for several days.  Dense polycrystalline material was produced using this method.

As explained in subsequent sections, some differences were noted in the behavior of the single
crystalline vs. the polycrystalline material.  Because the crystals were grown from the vapor, it is
likely that the stoichiometry of the crystals is slightly different from that of the polycrystalline
material.  To verify this, electron microprobe measurements were performed on both types of material,
and the ratio of the areas under the Cd $L$ line and the Os $M$ line were compared.  The measurements
indicated that the ratio of Cd to Os was slightly higher in the crystals.  Because a detailed
single-crystal refinement found no deviation from stoichiometry, it is likely that the
polycrystalline material is slightly Cd deficient, especially given that pyrochlores tolerate
significant A site vacancies \cite{Subramanian83}.  Small differences in oxygen content between the
single- and polycrystalline material also cannot be ruled out.

\section{Crystallography}

X-ray diffraction measurements on $\mathrm{Cd_2Os_2O_7}$ were performed  using a Nonius 4-circle
diffractometer equipped with Mo $Ka$ radiation and a nitrogen gas-stream cryo-cooler.  An equant
octahedral crystal, diameter 0.084 mm, was mounted on the diffractometer and nearly a full sphere of
reflections was collected at 295 K assuming no systematic absences.  Two smaller data sets were
collected at 250 K and 180 K. Because the crystal is highly absorbing, the diffraction data were
corrected for absorption analytically using the Gaussian face-indexed method.  The observed
systematic absences were consistent with an $F$-centered cell.  A combination of Patterson and
difference Fourier mapping confirmed the ideal pyrochlore structure type at all temperatures.

The refinement results appear in Table \ref{refine}.  The refined room temperature lattice parameter,
10.1651(4) {\AA}, is in good agreement with the 10.17 {\AA} reported in Ref. \onlinecite{Sleight74}.
The cell edge exhibits a smooth contraction with temperature and there is little or no change in the
structural parameters upon cooling.  The refined oxygen position, 0.319(2), is close to the value of
0.3125 that gives the maximum nearest-neighbor anion separation and regular BO$_{6}$ octahedra for
the ideal pyrochlore structure \cite{Chako84}.  The difference between the volume of the unit cell at
250 K and 180 K is ${\Delta}V/V$(250 K) = 0.047\%.  This volume change implies an average linear
thermal expansion coefficient (180-250 K) of a = 2.24 $\times 10^{-6} K^{-1}$ through the transition.
This is a small but not unreasonable value, and is consistent with the relatively high value of the
Debye temperature ($\Theta_D \approx$ 400 K) estimated below from specific heat measurements.  The
lattice parameter was measured at a few other temperatures in the vicinity of the transition, and no
anomalies were noted.  The picture that emerges from the crystallographic data is that of a stiff
lattice coupled only weakly to the electronic degrees of freedom.

\begin{table}
\caption{Crystal Data and Refinement$^{a}$ Results.} \squeezetable \label{refine}
\begin{ruledtabular}
\begin{tabular}{lccc}
 & 180 K  & 250 K  & 295 K\\ \hline
 a(\AA) & 10.1598(4) & 10.1614(4) & 10.1651(4)\\ $V($\AA$^{3})$ & 1048.7 & 1049.2 & 1050.1 \\
 measured reflections & 666 & 1271 & 3589 \\
 independent reflections & 139 & 139 & 139 \\
 reflections with $I > 4\sigma (I)$ & 113 & 114 & 113 \\
 $R_{int}$ & 0.048 & 0.049 & 0.050 \\
 $R(F)$ & 0.046 & 0.044 & 0.036 \\
 $wR(F^{2})$ & 0.224 & 0.305 & 0.305 \\
 goodness-of-fit, $S$ & 1.83 & 2.60 & 2.75 \\
 reflections in refinement & 139 & 139 & 139 \\
 parameters refined & 11 & 11 & 11 \\
 extinction coefficient & 0.0009(3) & 0.0009(4) & 0.0011(4) \\
 $x$(O1) & 0.319(1) & 0.319(2) & 0.319(2) \\
 O1 $U_{11}$ $($\AA$^2)$ & 0.010(6) & 0.009(8) & 0.007(7) \\
 O1 $U_{22} = U_{33}$ $($\AA$^2)$ & 0.008(3) & 0.011(5) & 0.012(5) \\
 O1 $U_{23}$ $($\AA$^2)$ & 0.000(5) & 0.002(8) & 0.002(8) \\
 O1 $U_{eq}$ $($\AA$^2)$ & 0.009(2) & 0.010(3) & 0.011(3) \\
 Cd $U_{11} = U_{22} = U_{33} = U_{eq}$ $($\AA$^2)$ & 0.008(5) & 0.009(1) &
 0.010(1) \\
 Cd  $U_{12} = U_{13} = U_{23}$ $($\AA$^2)$ & -0.006(4) & -0.0007(7) & -0.0009(5) \\
 Os $U_{11}= U_{22} = U_{33} = U_{eq}$ $($\AA$^2)$ & 0.0055(8) & 0.005(1) &
 0.005(1) \\
 Os  $U_{12} = U_{13} = U_{23}$ $($\AA$^2)$ & -0.0002(2) & -0.0001(3) & -0.0002(2) \\
 O2 $U_{11} = U_{22} = U_{33} = U_{eq}$ $($\AA$^2)$ & 0.008(6) & 0.007(8) & 0.012(9) \\
 Cd-O1 (\AA) $\times$ 6 & 2.57(1) & 2.57(1) & 2.57(1) \\
 Cd-O2 (\AA) $\times$ 2 & 2.1997(4) & 2.2000(4) & 2.2006(4) \\
 Os-O1 (\AA) $\times$ 6 & 1.926(5) & 1.926(8) & 1.928(7) \\
 Os-O1-Os ($^{\circ}$) & 137.5(7) & 137.5(7) & 137.5(7) \\
\end{tabular}
\end{ruledtabular}
\footnotetext[1]{ Refinement on ${F}$$^{2}$, weighting
 ${w}$ = 1/[${\sigma}$$^{2}$(${F}$$_{0}$$^{2}$) + (0.0558${P}$)$^{2}$ + 0.4843${P}$] where ${P}$ =
 0.1.
 ${U}_{ij}$ = exp[-2${\pi}^{2}{a}^{*2}$(${U}_{11}{h}^{2}+{U}_{22}{k}^{2}+{U}_{33}{l}^{2}+{2U}_{12}{hk}
 +{2U}_{13}{hl}+{2U}_{23}{kl}$)].  Atom positions: Cd ${16d}$ ${1/2, 1/2, 1/2}$; Os ${16c}$ ${0,0,0}$; O1
 ${48f}$ ${x, 1/8, 1/8}$; O2 ${8b}$ ${3/8, 3/8, 3/8}$.  Symmetry constraints on ${U}_{ij}$: Cd, Os  ${U}_{11}=
 {U}_{22}={U}_{33}$, ${-U}_{12}={-U}_{13}={-U}_{23}$; O1  ${U}_{11}$, ${U}_{22}={U}_{33}$, ${U}_{12}={U}_{13}=0$,
 ${U}_{23}$; O2 ${U}_{11}={U}_{22}={U}_{33}$, ${U}_{12}={U}_{13}={U}_{23}=0$.}
\end{table}

\section{Resistivity}

Resistance vs. temperature measurements were performed on both single crystals (SC) and
polycrystalline (PC) samples, and all specimens showed a well-defined MIT at approximately 226 K.
There was no indication of thermal hysteresis in the resistivity of either the PC or the SC material,
consistent with the continuous nature of the MIT in this material.  The measurements were performed
using a linear 4-probe method, 25 $\mu$m Pt wire, and Epotek H20-E silver epoxy. The resistivity of
two crystals appears in Fig.1 .  The room temperature resistivity was about 750 $\mu\Omega$ cm in
both crystals. When cooled from 226 K to 5 K, the resistivity increases by three orders of magnitude.
This large increase makes it likely that the entire Fermi surface has been eliminated. From 226 K to
750 K, the resistivity shows little temperature dependence. This behavior is in striking contrast to
the non-saturating, strongly temperature dependent behavior of the resistivity observed in oxides
such as the cuprates and in SrRuO$_3$ \cite{Allen96}. Given the extremely weak temperature dependence
observed in $\mathrm{Cd_2Os_2O_7}$, the mean free path of the carriers has evidently saturated on the
order of an interatomic spacing. The source of the strong scattering is evidently coulombic, and
likely involves exchange interactions within the $t_{2g}$ manifolds. Hund's rules favor parallel
spins within each manifold; however, for a half-full $t_{2g}$ manifold, the next electron should
enter with anti-parallel alignment.  Therefore, in an analogy with double exchange, the system will
tend toward antiferromagnetism so as to allow the electrons to hop more easily and gain kinetic
energy. However, at least for localized moments, the pyrochlore lattice is geometrically frustrated
\cite{Anderson56} \cite{Ramirez94}. Therefore, one possible source of the strong scattering lies in
the frustration inherent in the pyrochlore lattice. Reinforcing this notion is the fact that other
metallic $4d/5d$ pyrochlores, such as $\mathrm{Bi_2Ru_2O_7}$, also display resistivities that are
practically independent of temperature \cite{Pike77}.

\begin{figure}
\includegraphics[width = 3 in]{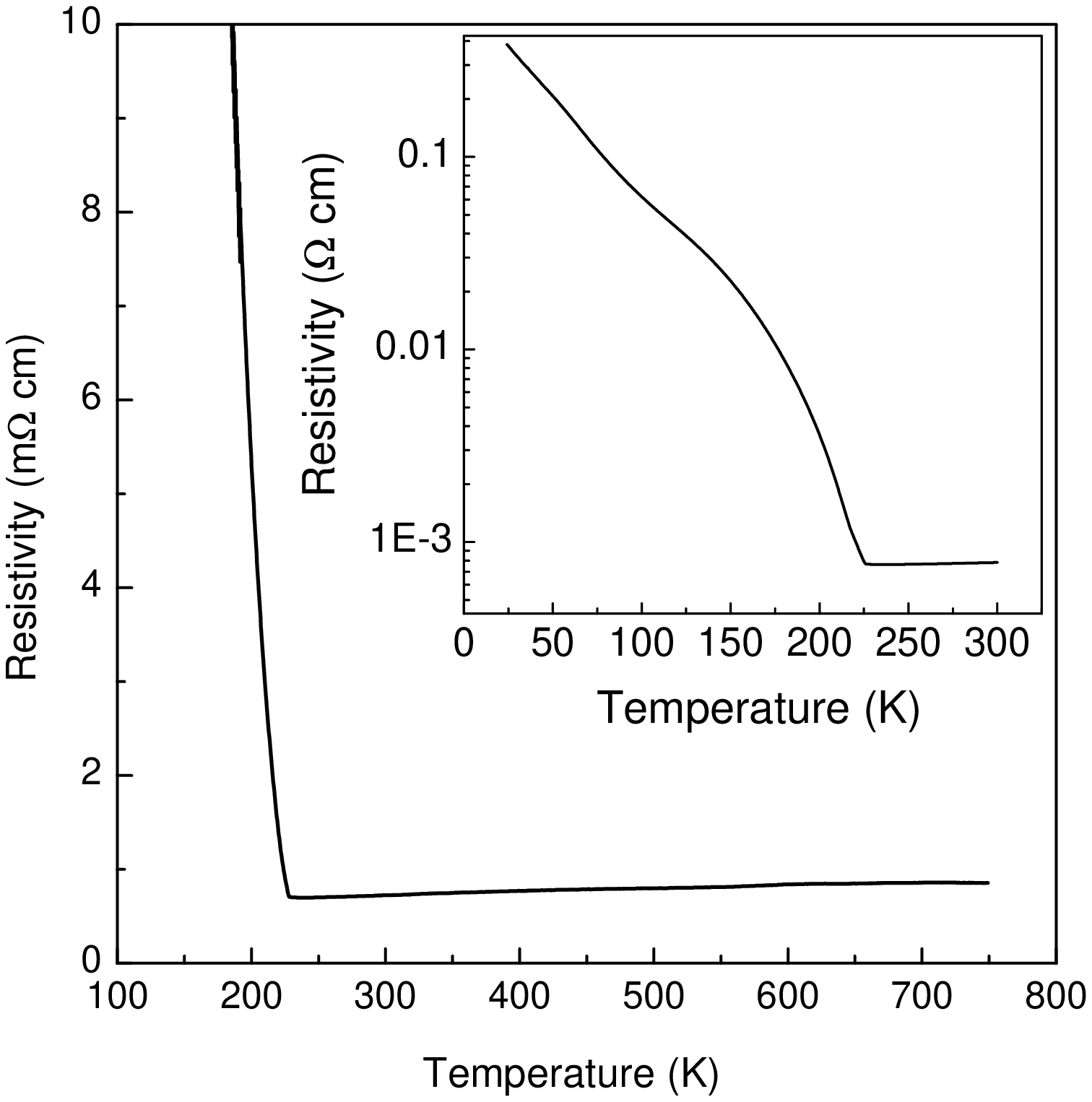}
\caption{Resistivity of two crystals of $\mathrm{Cd_2Os_2O_7}$.}
\end{figure}

\begin{figure}
\includegraphics[width = 3 in]{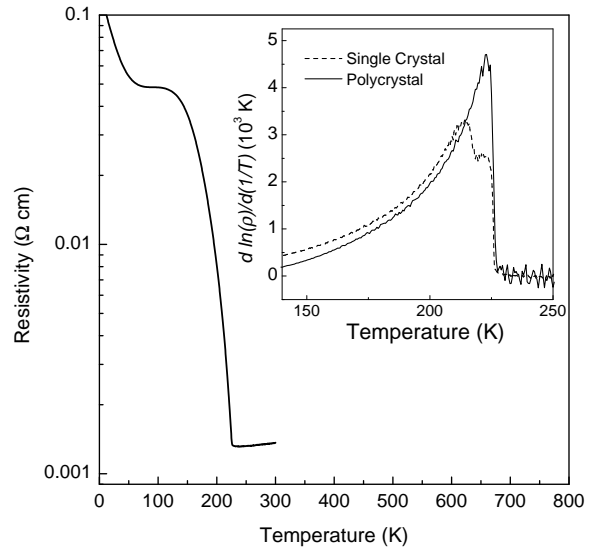}
\caption{Resistivity of a polycrystalline sample of $\mathrm{Cd_2Os_2O_7}$.  Inset: the quantity
$d(\ln \rho)/d(1/T)$ vs. temperature for a single crystal and a polycrystalline sample.  See text for
discussion.}
\end{figure}

\begin{figure}
\includegraphics[width = 3 in]{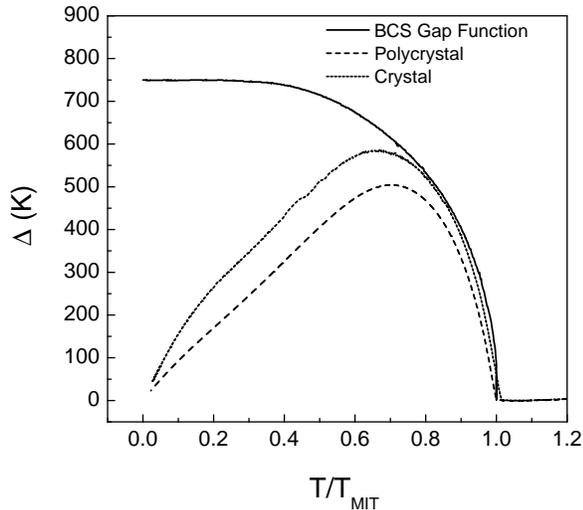}
\caption{Activation energy vs. temperature calculated for a single crystal and a polycrystalline
sample of $\mathrm{Cd_2Os_2O_7}$. The activation energy was calculated by assuming $\Delta = T \ln
{(\rho/\rho_0)}$ as described in the text.  A BCS gap function ($\Delta$(0) = 750 K) is also plotted
for comparison.}
\end{figure}

The resistivity of a PC sample of $\mathrm{Cd_2Os_2O_7}$ appears in Fig. 2. Qualitatively, the
overall temperature dependence of the resistivity of the PC material is quite similar to that of the
crystals, although the magnitude of the resistivity is of course higher in the PC material
($\rho$(295 K)= 1.3 m$\Omega$ cm). The change in resistivity upon cooling to 4 K is smaller in the PC
material as well, indicating a higher concentration of impurities in this material. In the inset to
Fig. 2 we plot the quantity $d(\ln \rho)/d(1/T)$ for both SC and PC samples.  This quantity has often
been used to emphasize the anomalous features in the resistivity of organic conductors and $1d$ SDW
materials \cite{Gruner94}.  The first point to notice is that although the change in the resistivity
itself is smooth, the change in the logarithmic derivative of the resistivity is abrupt and resembles
the specific heat anomaly (see Fig. 5).  This seems consistent with the continuous (second order)
nature of the phase transition.  In ferromagnetic metals, for example, the derivative of the
resistivity ${d\rho/dT}$ exhibits the same critical behavior as the specific heat \cite{Klein96}, and
something similar seems to be going on here.  The next point to notice is that the SC curve is
different from the PC curve in that it displays an additional feature with an onset near 217 K.  This
feature is robust and has been observed at the same temperature in the 6 crystals we have measured.
Furthermore, a feature at the same temperature has been observed in specific heat and magnetization
measurements on crystals as described below. Although chemical inhomogeneity is a possible
explanation of the 217 K feature, a number of reasons suggest that the feature may well be intrinsic.
First, the feature is reproducible in all the crystals measured.  Second, thinning the crystal to
${<}$ 50 ${\mu}$m did not affect the magnitude or shape of the resistivity. Third,  EDX spectra
obtained at a number of different places on a crystal showed no differences in the Cd/Os ratio.
Fourth, the x-ray refinement showed no evidence for chemical inhomogeneity.  Fifth, in the case of
chemical inhomogeneity, one would expect the second transition to appear as a pronounced increase in
the resistivity itself (at least on a log scale), and not just as a feature in the derivative. Sixth,
as discussed below, the magnetization data are more consistent with a homogeneous sample. Seventh,
the method used to grow the crystals---vapor transport---is slow, gentle, and well-suited to
producing homogeneous samples, although the exact stoichiometry is often difficult to control.
Lastly, it should be pointed out that the observation of a second transition in a higher quality
sample is not unprecedented.  In the case of EuB$_6$, for example, more than 30 years passed before
the discovery of the spin reorientation transition occurring a few degrees below $T_C$
\cite{Sullow98}.  It is possible that some sort of spin reorientation transition is occurring in
$\mathrm{Cd_2Os_2O_7}$ as well.

In a Slater picture of an MIT, we expect the resistivity to follow an expression of the form $\rho =
\rho_0e^{(\Delta/T)}$ as has been found for quasi 1d SDW materials \cite{Gruner94}. We also expect
that the temperature dependence of a Slater insulating  gap will behave in much the same way as a BCS
gap \cite{Matsubara54} \cite{DesCloizeaux59}.  In Fig. 3 we plot the temperature dependence of the
activation energy, $\Delta$, calculated from the experimental data using $\Delta = T \ln
{(\rho/\rho_0)}$. Here $\rho_0$ is the resistivity just above the transition.  Also in Figure 3 we
plot a BCS gap function ($\Delta$ = 750 K) for comparison. As is clear from the Figure, the
resistivity of $\mathrm{Cd_2Os_2O_7}$ is consistent with this picture at least for temperatures close
to the transition.  At lower temperatures the influence of extrinsic conduction mechanisms becomes
increasingly important and masks the intrinsic behavior.  This occurs at a higher temperature in the
PC material as compared to the SC material, consistent with the greater impurity concentration in the
PC material.  Assuming that $\Delta \approx$ 750 K, we find that $2\Delta \approx$ 6.6 $k_BT_C$. This
is considerably higher than the $2\Delta$ = 3.5 $k_BT_C$ predicted from weak coupling SDW theory,
but, interestingly, quite similar to the $2\Delta$/$k_BT_C$ values found in cuprate superconductors.

\section{Specific Heat}

The specific heat of SC and PC $\mathrm{Cd_2Os_2O_7}$ from 2 K to 300 K is plotted in Fig. 4. These
data were obtained using a commercial heat-pulse calorimeter manufactured by Quantum Design.  The
calorimeter is periodically tested against a sapphire standard to ensure reliable results.

A plot of $C_P/T$ vs. $T^2$ indicates that the lattice contribution above about 3.5 K is not
well-described by a simple Debye model.  This is not surprising because pyrochlores typically have
several low frequency optical phonons \cite{Subramanian83} \cite{McCauley73} which are expected to
strongly influence the low temperature heat capacity.  If we fit the data between 1.9 K and 3.5 K to
the low $T$ approximation $C_P=\gamma T + \beta T^3$, we find that for a sample consisting of about
10 single crystals (m = 18 mg) we have $\gamma$ = 1.08 mJ/mol-K$^2$ and $\Theta_D$ = 463 K and for a
polycrystalline disk (m = 49 mg) we have $\gamma$ = 1.4 mJ/mol-K$^2$ and $\Theta_D$ = 354 K, where
$\gamma$ is the Sommerfeld coefficient and $\Theta_D$ is Debye temperature in the limit
$T\rightarrow0$.  These $\gamma$ values are much smaller than those of related (metallic) pyrochlores
such as $\mathrm{Cd_2Re_2O_7}$ ($\gamma$ = 25 mJ/mol-K$^2$) \cite{Blacklock79}  and
$\mathrm{Cd_2Ru_2O_7}$ ($\gamma$ = 12 mJ/mol-K$^2$) \cite{Blacklock79} and suggest that the Fermi
surface is fully gapped below the MIT. This is consistent with the activation energy analysis of the
resistivity discussed in the previous section.

\begin{figure}
\includegraphics[width = 3 in]{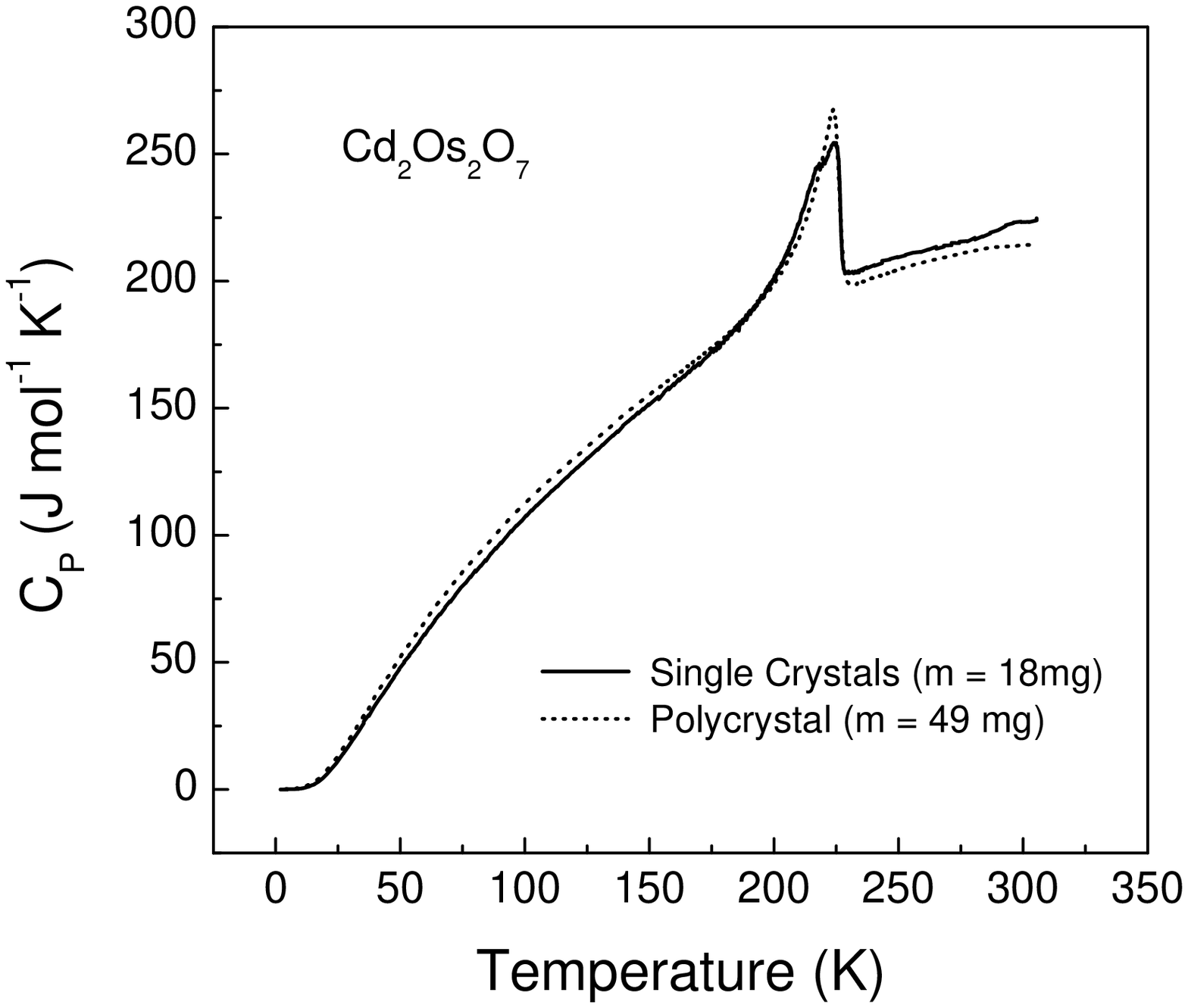}
\caption{Specific heat for single crystal and polycrystalline $\mathrm{Cd_2Os_2O_7}$.}
\end{figure}

\begin{figure}
\includegraphics[width = 3 in]{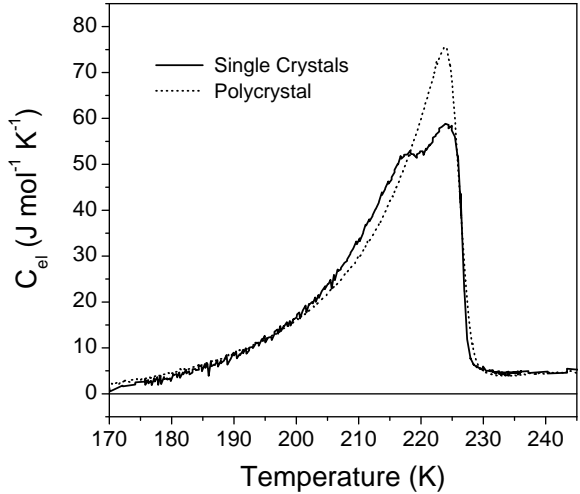}
\caption{The electronic portion of the specific heat obtained as described in the text.}
\end{figure}

The estimated electronic contribution to the specific heat is plotted in Fig. 5.  This estimate was
obtained by first \textit{assuming} a Sommerfeld coefficient above $T_{MIT}$ and subtracting off the
assumed electronic contribution.  Then a smooth polynomial was fitted to the data outside the region
of the anomaly as an estimate of the lattice contribution.  Then the lattice contribution was
subtracted from the raw data.  Sommerfeld coefficients between 0 and 30 mJ/mol-K$^2$ were explored,
but the entropy analysis described below is relatively insensitive to the precise value of the
assumed $\gamma$. A $\gamma$ of 20 mJ/mol-K$^2$ was used to produce Fig. 5.  This value is in the
range of the Sommerfeld coefficients reported on the metallic pyrochlores mentioned above and is
meant to provide a reasonable estimate of the entropy of the itinerant electrons at temperatures
above the MIT.

The shape of the specific heat anomaly resembles a mean-field, BCS-type transition.  In single
crystals there are two features in the specific heat analogous to the features in the ${d(\ln
\rho)/d(1/T)}$ plots discussed earlier.  No hysteresis was observed in the specific heat
measurements; the data taken upon cooling or warming, were, within the scatter, indistinguishable.
Given that there is no hysteresis, no latent heat, and a volume change of less than 0.05\%, we
conclude that the MIT in $\mathrm{Cd_2Os_2O_7}$ is continuous.

Let us now consider how electronic entropy is eliminated as $T\rightarrow0$.  A system of localized
$d$ electrons has a large electronic entropy of order $k_B\ln{N}$ per magnetic ion, where $N$ is the
degeneracy of the ground state of the atomic $d$ electrons subject to Hund's rule and crystal field
interactions. In this case, entropy is typically eliminated as $T\rightarrow0$ by a transition to a
magnetically ordered state. In a system of itinerant electrons, on the other hand, the specific heat
is given by $C = \gamma T$, and the Pauli principle ensures that the entropy vanishes as
$T\rightarrow0$.  In $\mathrm{Cd_2Os_2O_7}$, we can estimate the entropy associated with the MIT by
integrating $C_{el}/T$ from 170 K to 230 K.  Depending on the assumed value of $\gamma$ above
$T_{MIT}$, the answer ranges (in single crystals) from $S_{MIT}$ = 5.4 J/mol-K ($\gamma=0$
mJ/mol-K$^2$) to $S_{MIT}$ = 6.6 J/mol-K ($\gamma=30$ mJ/mol-K$^2$) and (in polycrystals) from
$S_{MIT}$ = 5.6 J/mol-K ($\gamma=0$ mJ/mol-K$^2$) to $S_{MIT}$ = 6.8 J/mol-K ($\gamma=30$
mJ/mol-K$^2$).  For an assumed $\gamma$ of 20 mJ/mol-K$^2$ we have $S_{MIT}$ = 6.2 J/mol-K (SC) and
$S_{MIT}$ = 6.5 J/mol-K (PC).

Localized Os$^{5+}$($5d^3$) ions are expected to have spin-3/2 and to eliminate $2R\ln{4}$ = 23.0
J/mol-K via a magnetic transition. This is clearly much higher than the observed value, and suggests
that the transition does not involve ordering of localized $5d$ moments.  Even if we assume that
spin-orbit coupling breaks the degeneracy of the $t_{2g}$ manifold making the ions effectively
spin-1/2, we still expect an entropy of $2R\ln{2}$ = 11.5 J/mol-K which is again significantly higher
than the observed value.  It seems more reasonable to identify S$_{MIT}$ with the entropy of an
itinerant electron system above the MIT.  If we make this association, S(230 K) = 4.6 J/mol-K for
$\gamma=20$ mJ/mol-K$^2$ and S(230 K) = 6.9 J/mol-K for $\gamma=30$ mJ/mol-K$^2$.  These numbers are
much closer to the observed values and support the notion that the entropy associated with the MIT in
$\mathrm{Cd_2Os_2O_7}$ is simply that of the itinerant electron system above T$_{MIT}$.  This picture
is consistent with what we expect from a Slater transition.

\section{Magnetization}

\begin{figure}
\includegraphics[width = 3 in]{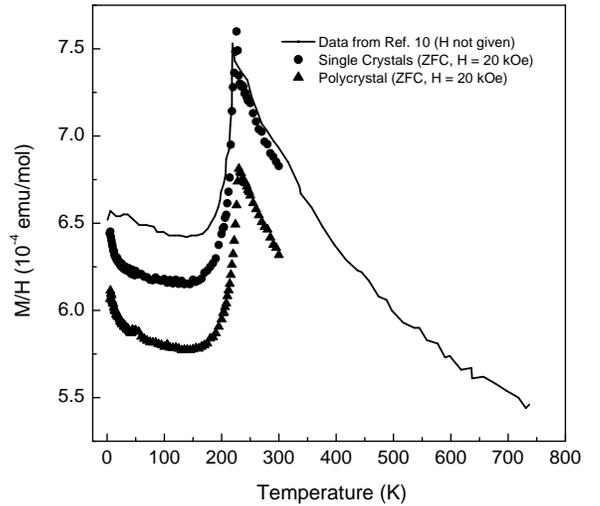}
\caption{Effective susceptibility ($M/H$) obtained on single and polycrystalline
$\mathrm{Cd_2Os_2O_7}$ under zero field cooled conditions with an applied field of 20 kOe. Data from
Ref. \onlinecite{Sleight74} obtained under unspecified conditions are also plotted.}
\end{figure}

\begin{figure}
\includegraphics[width = 3 in]{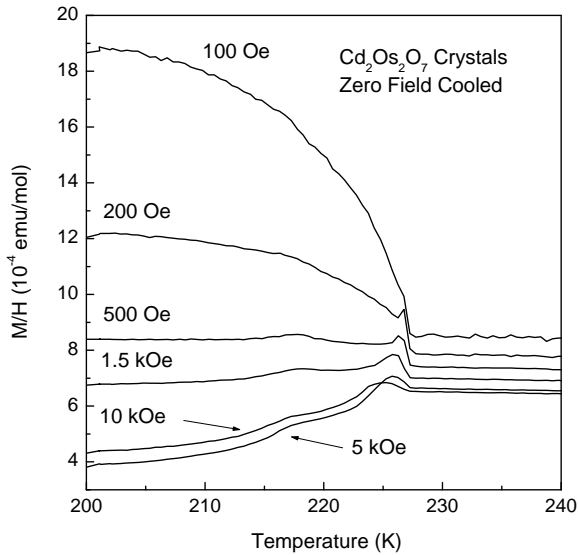}
\caption{$M/H$ vs. temperature for single crystals of $\mathrm{Cd_2Os_2O_7}$ obtained under zero
field cooled conditions with applied fields as indicated.  Note the additional structure at T = 217
K, corresponding to features in the resistivity and heat capacity.}
\end{figure}

\begin{figure}
\includegraphics[width = 3 in]{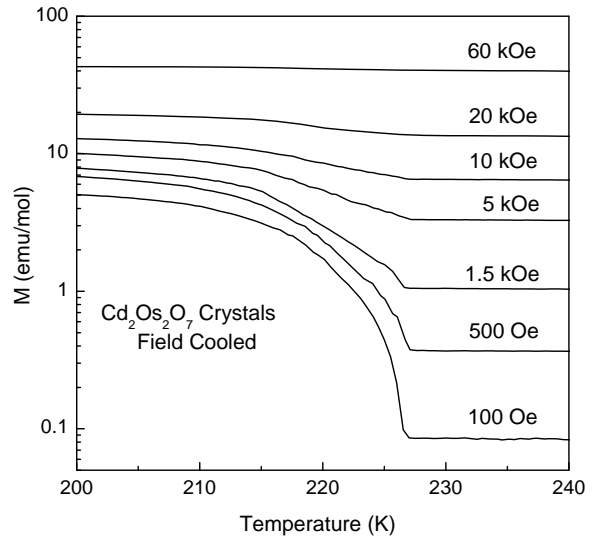}
\caption{Magnetization vs. temperature for single crystals of $\mathrm{Cd_2Os_2O_7}$ obtained under
field cooled conditions with applied fields as indicated.}
\end{figure}

\begin{figure}
\includegraphics[width = 3 in]{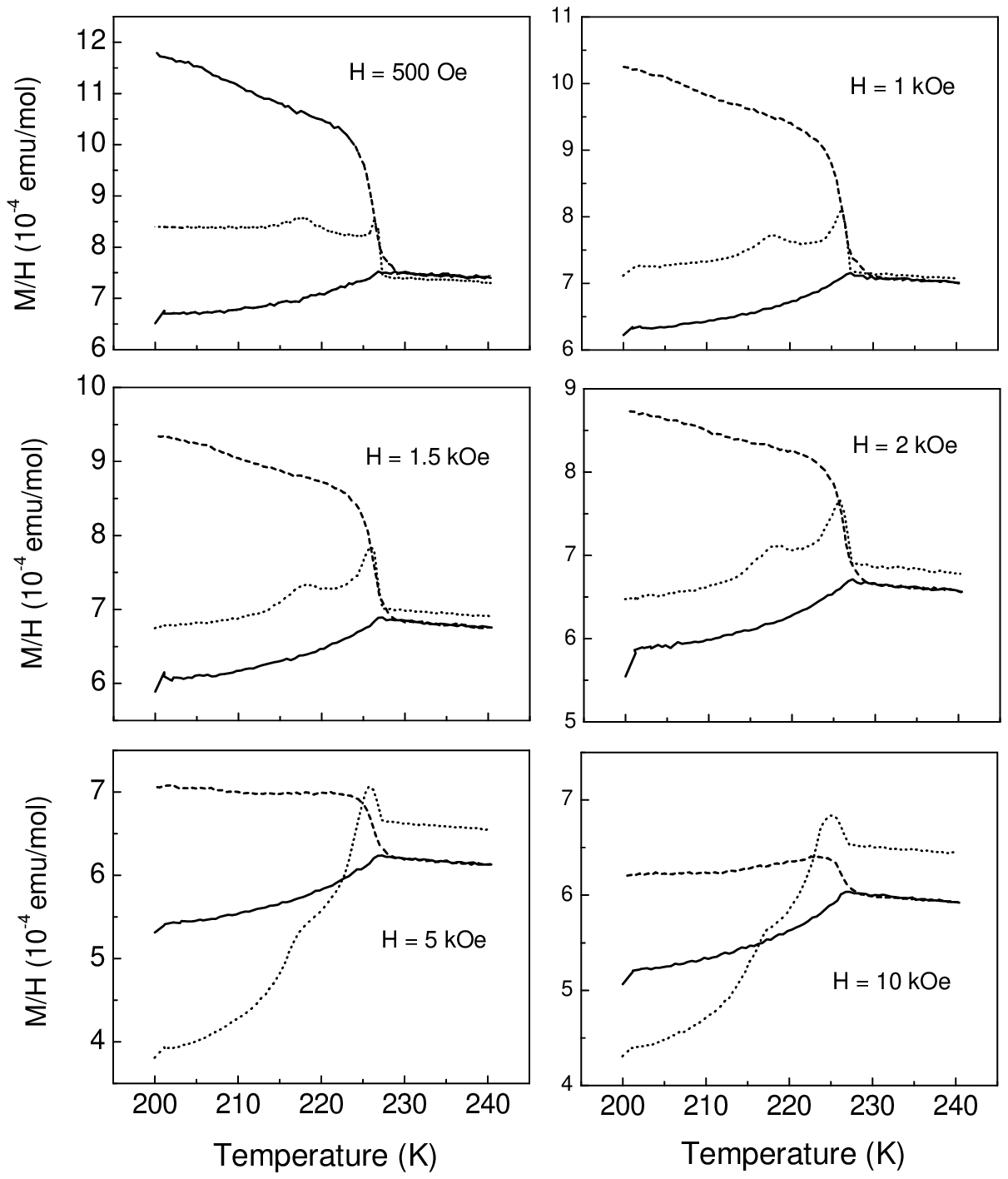}
\caption{$M/H$ vs. temperature for polycrystalline $\mathrm{Cd_2Os_2O_7}$ with applied fields as
indicated. Both ZFC (solid line) and FC (dashed line) are shown.  The dotted line is ZFC single
crystal data included for comparison.}
\end{figure}

Magnetization measurements were performed on $\mathrm{Cd_2Os_2O_7}$ using a SQUID magnetometer from
Quantum Design. The measurements were performed by cooling the sample in zero field, turning on the
magnet, and then measuring the moment as the sample was warmed (ZFC) and then re-cooled (FC).  In
Fig. 6 we plot $M/H$ vs. T (ZFC, H = 20 kOe) for SC and PC samples and for comparison we plot the
magnetic susceptibility data obtained on single crystals from Ref. \onlinecite{Sleight74}.  The SC
results are in good quantitative agreement.  The magnetic response of the PC material is somewhat
weaker, but qualitatively the agreement is good.  According to Ref. \onlinecite{Sleight74}, the data
from 226-750 K do not obey a Curie-Weiss law.  If, as we concluded from the specific heat analysis,
the system above $T_{MIT}$ is a moderately correlated metal, we do not expect the magnetic response
to obey a Curie-Weiss law.  What we expect is an exchange-enhanced Pauli paramagnetic response, and
the data are fully consistent with that.

In Figs. 7 and 8 we present ZFC and FC results obtained on SC samples for a number of applied fields.
In Fig. 9 we present similar data for the PC material, with the SC ZFC results included for
comparison.  The magnetic interactions appear to be predominantly antiferromagnetic, although a
parasitic ferromagnetic component is clearly present. Parasitic ferromagnetism is observed in many
antiferromagnetic materials (including SDW materials like $\alpha$-Mn) and is characterized by small
moments and by hysteresis with high coercive forces. Parasitic ferromagnetism is produced by defects
in an antiferromagnetic background becoming ``frozen-in''.  A good phenomenological account of
parasitic ferromagnetism is given by Arrott \cite{Arrott66}.  Isothermal magnetization curves $M(H)$
were reversible at all temperatures, except for a narrow interval just below the MIT.

At 5 K and 1000 Oe, the ZFC net moment is 1.0 emu/mol = $1.8 \times 10^{-4} \mu_B$ per formula unit;
the corresponding FC moment is 12 times larger.  The field dependence of the ZFC magnetization is
unusual as shown in Figs. 7 and 9. Note that the SC data obtained in magnetic fields between 500 Oe
and 10 kOe show clear evidence of a well-defined transition at 227 K and a broader transition near
217 K. These features correlate with similar features observed in the resistivity and specific heat
as discussed earlier.   It is interesting that at higher applied fields the 217 K feature becomes
weaker and is not visible in the 20 kOe data shown in Fig. 6. The magnetization data provides further
evidence that the additional feature at 217 K in the single crystals is likely to be intrinsic.  If
the SC samples were chemically inhomogeneous, with some portions ordering at 226 K and other portions
ordering at 217 K, then one would expect to see an additional feature near 217 K in the FC data in
Fig. 8.  The lack of a feature near 217 K in the FC data is good evidence that the entire sample has
ordered antiferromagnetically at 226 K.

It is difficult to imagine that $\mathrm{Cd_2Os_2O_7}$ is simply a local-moment antiferromagnet. In
the first place, magnetically ordered Os compounds are extremely rare and $4d/5d$ materials in
general tend to have low ordering temperatures.  Secondly, the pyrochlore lattice is known to be
geometrically frustrated \cite{Ramirez94} \cite{Anderson56} and for antiferromagnetic
nearest-neighbor interactions no long range order is predicted in the absence of further-neighbor
interactions \cite{Reimers91}.  In fact, antiferromagnetism is quite rare in pyrochlores
\cite{Subramanian83}. Given all this, a N\'{e}el temperature of 226 K is remarkably high and demands
explanation.

\section{Electrical and Thermal Transport}

\begin{figure}
\includegraphics[width = 3 in]{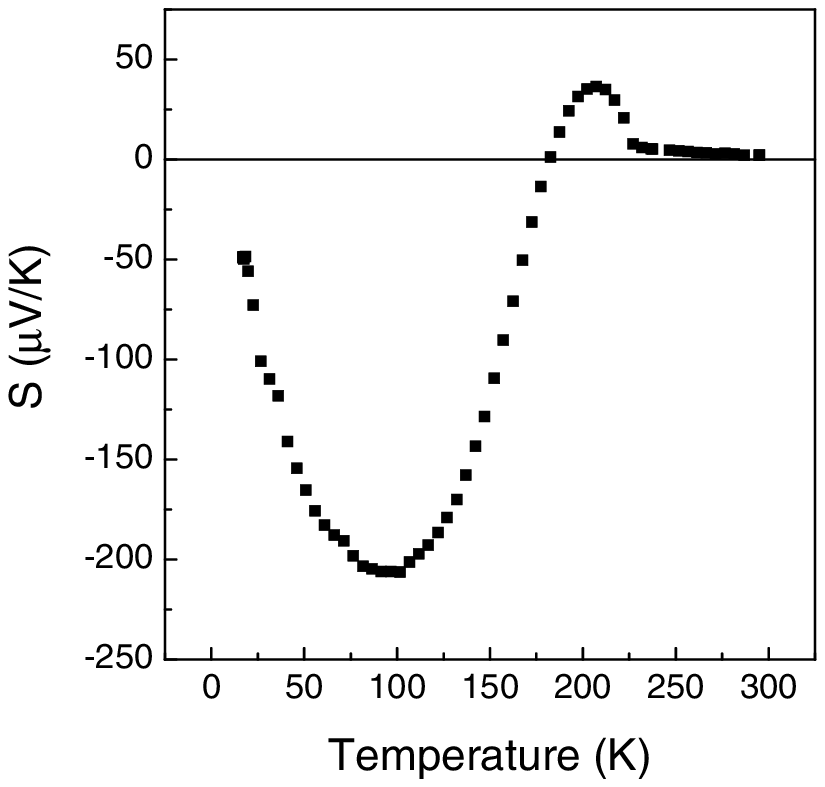}
\caption{Seebeck coefficient vs. temperature for a polycrystalline sample of $\mathrm{Cd_2Os_2O_7}$.}
\end{figure}

\begin{figure}
\includegraphics[width = 3 in]{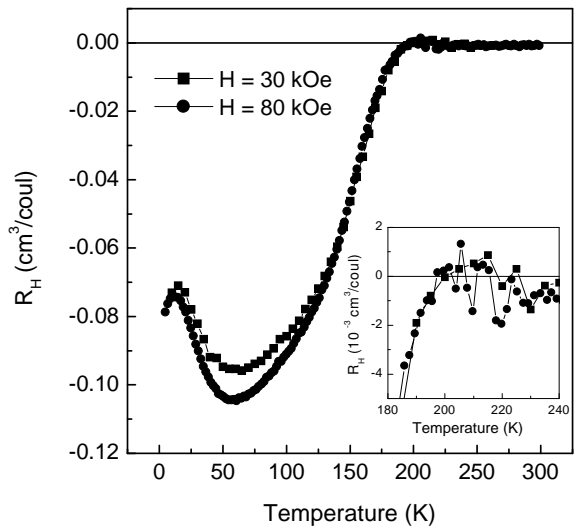}
\caption{Hall coefficient vs. temperature for a polycrystalline sample of $\mathrm{Cd_2Os_2O_7}$ at
applied fields of 30 kOe and 80 kOe as indicated.  The inset shows the behavior of the data near the
transition on an expanded scale.}
\end{figure}

Thermopower, Hall, and thermal conductivity measurements were performed only on PC samples due to the
difficulty of performing these measurements on extremely small crystals. For the thermopower and
thermal conductivity experiments the sample was a dense rod, 20 mm long and 8 mm in diameter. The
apparatus used for the thermopower and thermal conductivity measurements has been described
previously \cite{sales97}. This apparatus has been used extensively in thermoelectrics research and
has been tested repeatedly against silica and stainless steel standards. The Seebeck data were
corrected for the absolute thermopower of the Cu leads.  Hall measurements were performed in a
4-probe geometry at 30 kOe and 80 kOe. At each temperature, the sample was rotated 180$^{\circ}$ and
the current reversed in order to eliminate misalignment and thermal voltages, respectively. Checks
were made at several temperatures to ensure that the Hall voltage was linear in current.

In Fig. 10 we plot the Seebeck coefficient of $\mathrm{Cd_2Os_2O_7}$ vs. temperature.  Above
$T_{MIT}$, the thermopower is small and positive.  As the sample is cooled, the thermopower increases
from about 2 $\mu$V/K at 295 K to 6 $\mu$V/K at 230 K.  Such small Seebeck coefficients are
characteristic of metals.  As the sample is cooled below $T_{MIT}$, the behavior of the thermopower
changes drastically, first increasing to +40 $\mu$V/K near 200 K, and then changing sign and
decreasing to -225 $\mu$V/K near 90 K.  Although this behavior is complicated, it is clearly
consistent with a gap opening at the Fermi energy and a change from metallic behavior above $T_{MIT}$
to semiconducting behavior below $T_{MIT}$.

The Hall coefficient of $\mathrm{Cd_2Os_2O_7}$ appears in Fig. 11.  Above $T_{MIT}$ the magnitude of
$R_H$ is small as expected for a metal.  The differing signs of the Hall and Seebeck coefficients
indicate that both electrons and holes are participating in the electrical transport and preclude a
simple one-band analysis of the data.  Even with contributions from two types of carrier, however,
the Hall number can still give a useful order-of-magnitude estimate of the carrier concentration.  In
this spirit, we find that the room temperature Hall number is $n_H=8\times10^{21}$ cm$^{-3}$ and the
room temperature Hall mobility is $\mu_H$ = 0.6 cm$^2$ V$^{-1}$ s$^{-1}$.  These values are similar
to other metallic oxides such as $\mathrm{Fe_3O_4}$ and are not wildly different from cuprate
superconductors \cite{Tsuda91}.

Interestingly, no dramatic change in the Hall coefficient occurs until the sample is cooled to 200
K---well below $T_{MIT}$.  This is understandable if one imagines that a gap is opening as
illustrated in Fig. 3. When $\Delta \leq k_BT_C$ then thermal energy ensures that both bands are
equally populated and the carrier concentration hardly changes; it is only when $\Delta > k_BT_C$
that the carrier concentration begins to fall and the magnitude of the Hall coefficient begins to
increase.  The behavior of the thermopower is also understandable at a phenomenological level.  As
explained pedagogically by Chaikin \cite{Chaikin90}, the thermopower is a measure of the heat carried
by the electrons and/or holes.  As soon as a gap opens in the excitation spectrum, the thermopower of
both the electrons and holes increases because the carriers are now transporting the additional
energy required to create an electron-hole pair. The measured thermopower is a weighted average of
the electron and hole thermopowers, but if both electron and hole thermopowers increase in magnitude
it stands to reason that the weighted average is also likely to increase as the cancellation between
the electron and hole contributions is unlikely to be exact.

\begin{figure}
\includegraphics[width = 3 in]{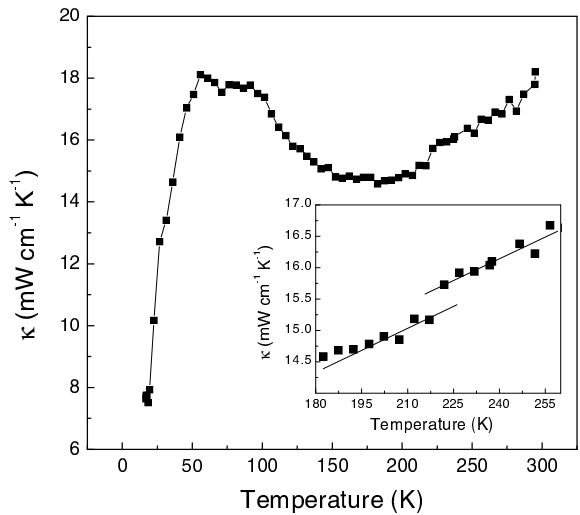}
\caption{Thermal conductivity vs. temperature for a polycrystalline sample of $\mathrm{Cd_2Os_2O_7}$.
The inset shows the behavior of the data near the transition on an expanded scale.  The lines through
the data are simply guides to the eye.}
\end{figure}

\begin{figure}
\includegraphics[width = 3 in]{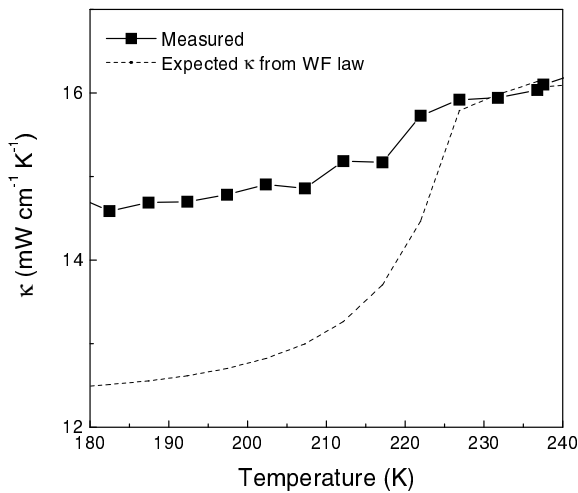}
\caption{Thermal conductivity of a polycrystalline sample of $\mathrm{Cd_2Os_2O_7}$.  The dashed line
is the expected change in thermal conductivity based upon the assumption that the Wiedemann-Franz law
is valid and that the Lorentz number takes its nominal value of $L_0=2.44\times10^{-8}$
W$\Omega$/K$^2$.}
\end{figure}

The thermal conductivity of a polycrystalline sample of $\mathrm{Cd_2Os_2O_7}$ appears in Fig. 12.
Below 150 K, the thermal conductivity resembles that of an ordinary crystalline solid---rising from
zero at low temperature, peaking near 50 K, and falling off as $1/T$ at higher temperatures. However,
near 150 K the thermal conductivity begins rising again in an unusual way.  Although radiation losses
can produce these high temperature ``tails'' in the thermal conductivity of low $\kappa$ materials,
is it unlikely that radiation losses are causing the rise here because our apparatus has been
optimized for use on low thermal conductivity samples. For example, silica samples roughly the same
size as the sample of $\mathrm{Cd_2Os_2O_7}$ required no radiation correction in our apparatus even
though the thermal conductivity of silica is roughly half that of $\mathrm{Cd_2Os_2O_7}$.  A more
likely explanation of the data can be understood from the simplest expression of the lattice thermal
conductivity of a solid which is given by $\kappa_L=(1/3)C_vv_sd$, where $C_v$ is the heat capacity
per unit volume, $v_s$ is an average sound velocity, and $d$ is the mean free path. What is likely to
be happening in $\mathrm{Cd_2Os_2O_7}$ is that the mean free path of the phonons does not decrease
any more above 150 K due to strong scattering, and that the rise in thermal conductivity simply
reflects the heat capacity which is still increasing due to the relatively high Debye temperature of
this material.

The MIT is reflected in the thermal conductivity as shown in the inset to Fig. 12.  The effect is
remarkably small considering the large changes in the electronic structure that are occurring at the
transition.  To quantify things, we plot in Fig. 13 the change in thermal conductivity we would
expect if the electronic portion of the thermal conductivity $\kappa_{el}$ were given by the
Wiedemann-Franz (WF) law.  The WF law states that the ratio between $\kappa_{el}$ and $\sigma$ is
given by $\kappa_{el}/\sigma = TL_0$ where $L_0=2.44\times10^{-8}$ W$\Omega$/K$^2$.  To produce Fig.
13 we used the WF law to estimate the electronic contribution to the thermal conductivity above
$T_{MIT}$, and then used the WF law and the measured resistivity to predict the change in thermal
conductivity we would expect if the WF law were obeyed in $\mathrm{Cd_2Os_2O_7}$.  As is clear from
the Figure, the predicted change is much greater than the measured change.  There are three possible
reasons for this discrepancy, and all are interesting.  The first is that the lattice contribution to
the thermal conductivity $\kappa_L$ increases as the sample is cooled below $T_{MIT}$. This
possibility is consistent with the overall behavior of the thermal conductivity discussed above. In
this scenario the phonons are strongly scattered by some sort of electronic excitations above
$T_{MIT}$, but below the transition these excitations disappear and the mean free path of the phonons
increases. Elastic modulus measurements can shed some light on the viability of this picture, and
these measurements are planned for the near future.  The second possibility is that magnetic
excitations are contributing to the thermal conductivity below $T_{MIT}$.  Measurements of the
magnetic structure and dynamics would be useful here. Lastly, the third possibility is that the WF
law is simply not obeyed in $\mathrm{Cd_2Os_2O_7}$. The WF law is empirical and is not perfectly
obeyed in real materials, although deviations as large as observed here are not common. Calculations
by Schultz and Allen \cite{Schultz95} show that as the scattering mechanism changes the Lorentz
number can vary between 0 (pure Coulomb scattering) and $L_0$ (pure electron-phonon scattering).  The
Lorentz number we infer from the analysis above is $L \approx 0.2 L_0$. Further study is required to
resolve these matters.

\section{Pressure dependence}

\begin{figure}
\includegraphics[width = 3 in]{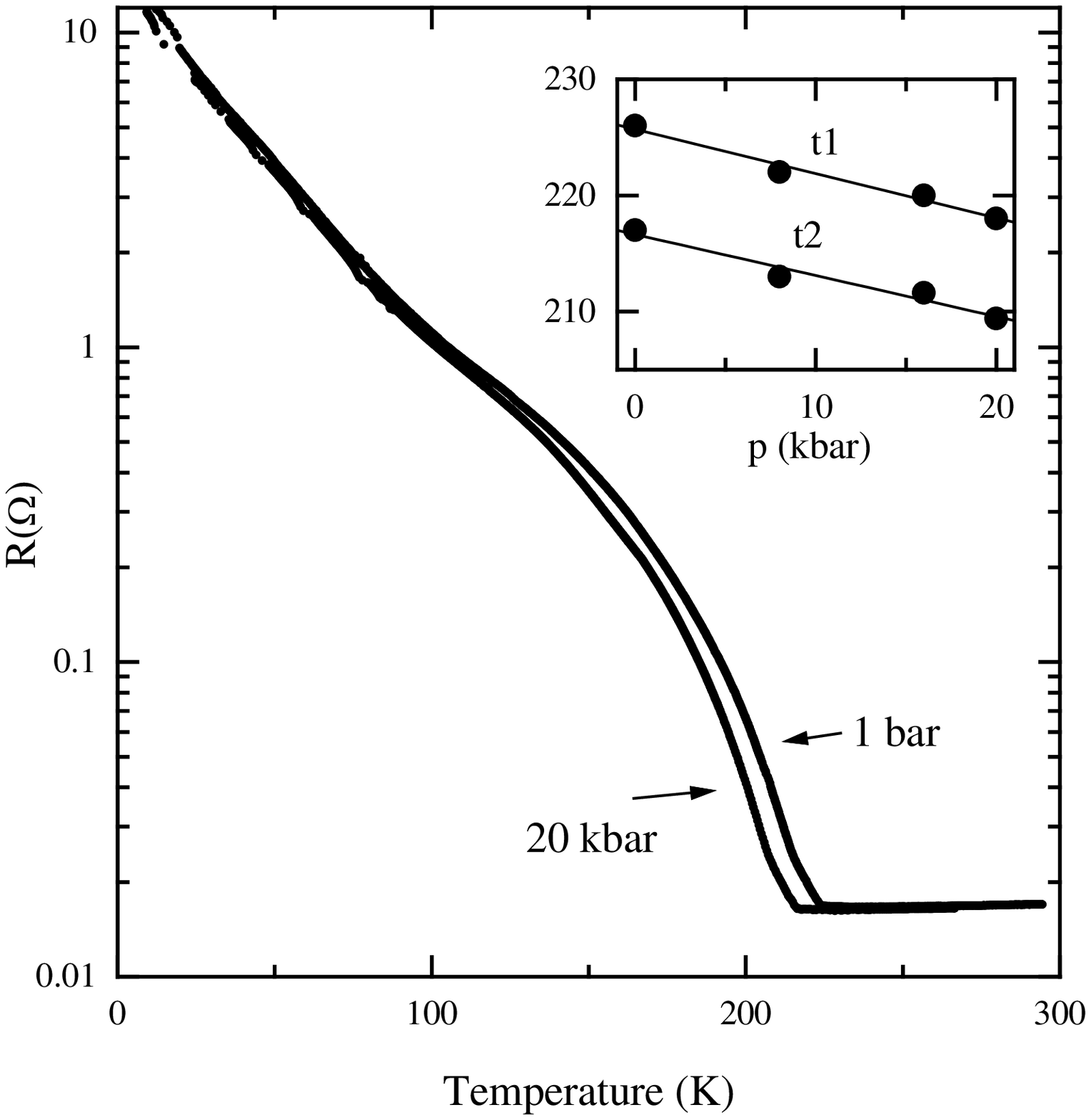}
\caption{R vs. T curves of a Cd$_{2}$Os$_{2}$O$_{7}$ single crystal at ambient pressure and at 20
kbar. The inset shows the two phase transition temperatures as a function of pressure.}
\label{1_20kbar_rt}
\end{figure}

In the case of the Mott-Hubbard transition, the pressure dependence has helped to understand the
underlying physics. To study further the nature of the phase transition in Cd$_{2}$Os$_{2}$O$_{7}$,
we have measured the temperature dependence of the resistivity of a single crystal under hydrostatic
pressures up to 20 kbar. The pressure cell was a self-clamping piston-cylinder cell using kerosene as
pressure medium. An InSb pressure sensor was mounted next to the sample, and pressure was monitored
during cooling. Typical pressure drop was 1 to 2 kbar during cool-down.

In Figure~\ref{1_20kbar_rt} we show the temperature dependence of the resistivity at ambient pressure
and at 20 kbar. The two curves virtually lie on top of one other, except between about 150 K and the
MIT transition region. The two transitions shift to lower temperatures at the same rate of about -0.4
K/kbar. This is twenty times smaller than the pressure-induced shift of the Mott-transition in pure
V$_{2}$O$_{3}$ \cite{V2O3carter}. We have also measured the pressure dependence of the room
temperature resistivity, which is small, -0.18\%/kbar (not shown).

These features suggest a rigid lattice to which electrons couple only weakly. In the case of
V$_{2}$O$_{3}$, the strong pressure dependence of the MIT is due to about 1\% volume change at the
transition. In our material no change beyond thermal expansion in the unit cell volume was detected
at the transition temperature thus the much weaker pressure dependence is natural. Since the Slater
transition is not critically sensitive to bandwidth as opposed to the Mott transition, this result is
not surprising. In V$_{2}$O$_{3}$, pressure also changes the activation energy in the insulating
state since for a narrow band the bandwidth increases exponentially with pressure, and the
Mott-Hubbard gap decreases linearly with the bandwidth, $\Delta=U-W$. Performing the same analysis as
described in the previous section, we did not find any change in the gap.

\section{Electronic Structure}

\begin{figure}
\includegraphics[width = 3 in]{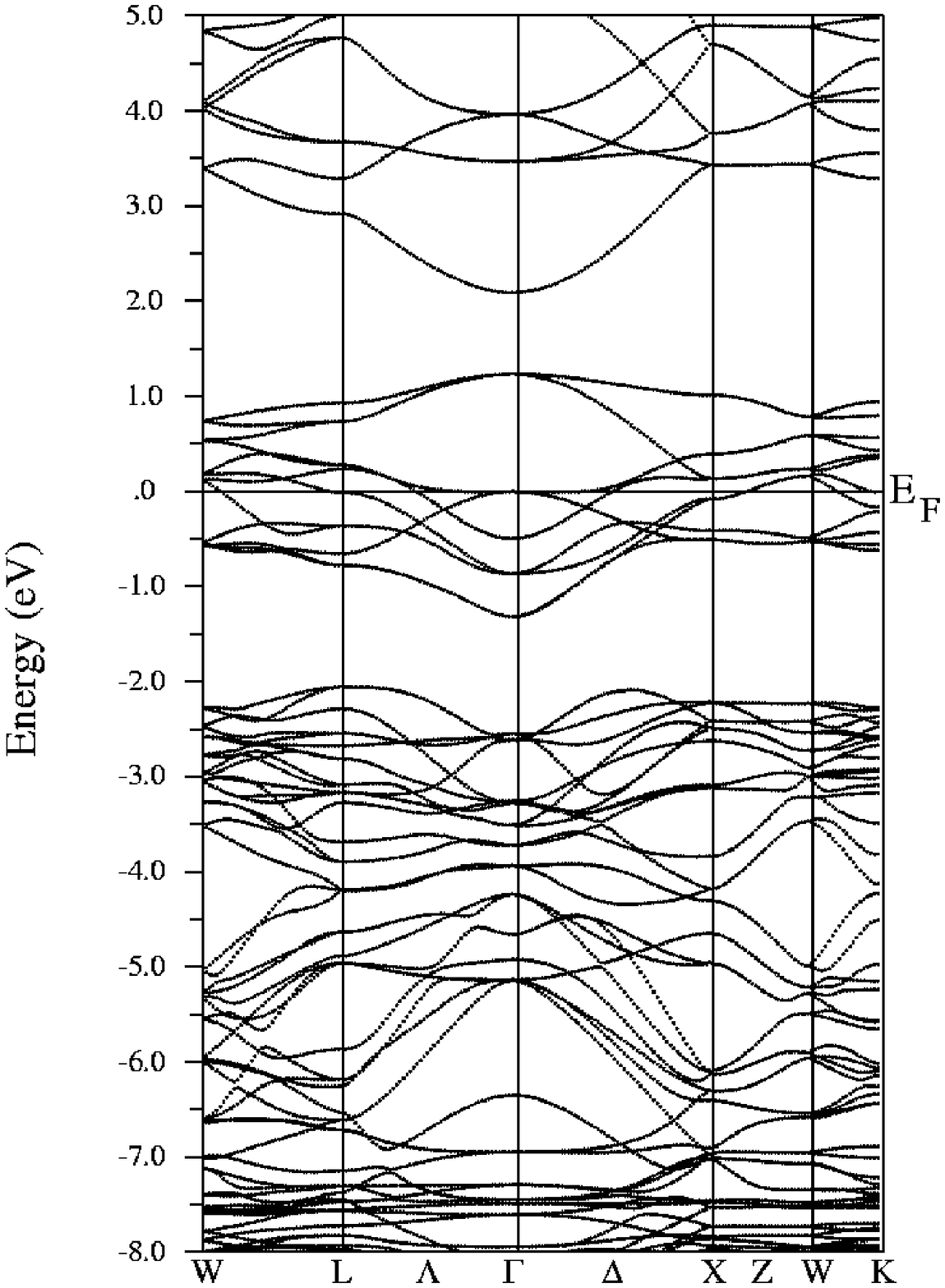}
\caption{Band structure of the pyrochlore $\mathrm{Cd_{2}Os_{2}O_{7}}$.}
\end{figure}

\begin{figure}
\includegraphics[totalheight = 3 in, width = 3 in]{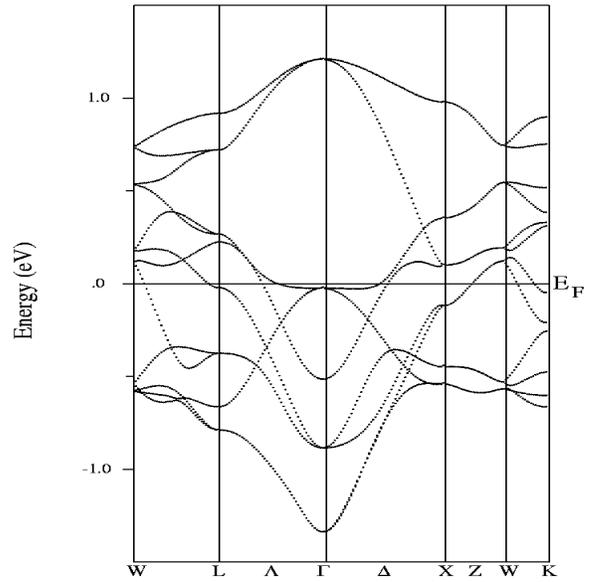}
\caption{Blowup of  the band structure of Fig. 15 near the Fermi energy.}
\end{figure}

\begin{figure}
\includegraphics[width = 3 in]{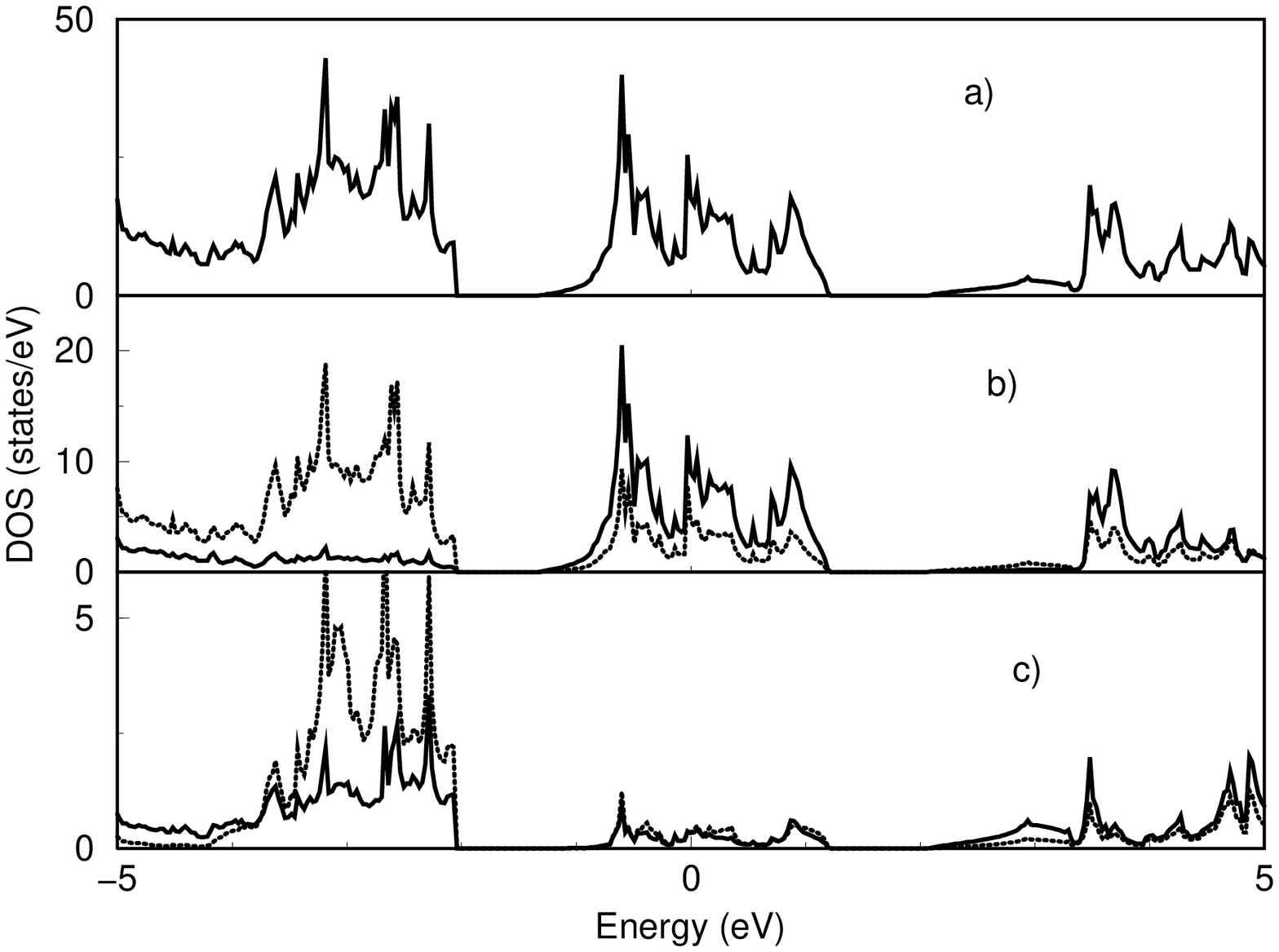}
\caption{Electronic DOS and projections onto LAPW spheres for $\mathrm{Cd_{2}Os_{2}O_{7}}$.  The
projections are on a per atom basis, while the total DOS is per unit cell.  The primitive cell used
in the calculations contains 2 formula units. a) Total DOS of $\mathrm{Cd_{2}Os_{2}O_{7}}$; b)
projection of Os (solid line) and O(1) (dotted line); c) projection of Cd (solid line) and O(2)
(dotted line).}
\end{figure}

The full-potential linearized augmented plane wave (LAPW) method in the WIEN97 code \cite{blaha} was
used for the electronic structure calculations.  In this method the unit cell is partitioned into
spheres (with a muffin-tin radius $R_{mt}$) centered on the atomic positions.  This method is an
application of density-functional theory in which the Kohn-Sham equations are solved in a basis of
linearized augmented plane waves with local orbital extensions. For the exchange and correlation
potential the generalized gradient approximation (GGA) of Perdew-Burke-Ernzerhof was used
\cite{perdew}. The core electrons are treated relativistically, whereas the valence electrons are
treated with a scalar non-relativistic procedure. The space group for this cubic pyrochlore compound,
$Fd\overline{3}m$, is face-centered and it requires four inequivalent atoms --- Cd, Os, O(1) and
O(2). The calculation we report here is non-spin-polarized (not magnetically ordered) and the
structural parameters from Table I are used. For the Cd the states up to $4s$ are taken as core
states, for Os the core states are up to $5p$, and for the O $1s$ is taken to be the core state. The
plane-wave cutoff is $R_{mt}K_{max}=7.0$ and we used 56 $k$ points in the irreducible wedge of the
Brillouin zone.

The band structure of $\mathrm{Cd_{2}Os_{2}O_{7}}$ is presented in Figs. 15 and 16.  A flat band
around the $\Gamma$-point crosses $E_{F}$; therefore, the non-spin-polarized calculation predicts
that this material is a metal. This is, of course, precisely what we expect above $T_{MIT}$ in a
Slater scenario.  The states near the Fermi energy have mainly Os $t_{2g}$ character, but there is
significant O $2p$ character as well. The unoccupied Os $e_{g}$ band is located roughly 0.8 eV above
the $t_{2g}$ band.  We also performed a similar calculation with the addition of spin-orbit coupling.
We find that the band structure in this case is the same as in Fig. 15, but the bands shift upwards
slightly by 0.08 eV.

Now we examine the corresponding density of states (DOS).  The total electronic density of states
(DOS) is presented in Fig. 17a and Fig. 18.  A sharp peak at the Fermi energy reaches $N(E_F)$ = 25.4
states/eV and implies a Sommerfeld coefficient $\gamma$ = 29.9 mJ/mol-$K^2$. Interestingly, this
value is in good agreement with the estimates made in Section VI.  Fig. 17b presents the partial DOS
of the Os and O(1). The Os-O(1) distance of 1.93 {\AA} is small enough to allow significant mixing of
the Os $t_{2g}$ and O $2p$ states. From Fig. 17b the hybridization between the two is evident.
Finally, Fig. 16c displays the partial DOS of the Cd and O(2). Hybridization here is also present,
and these states have a presence at $E_{F}$. A similar effect was noticed in another pyrochlore
material, $\mathrm{Tl_{2}Mn_{2}O_{7}}$, where the hybridization not only between Mn and one of the O
was found, but also between Tl and the other O \cite{singh}. In $\mathrm{Tl_{2}Mn_{2}O_{7}}$, the
hybridized Tl-O states needed to be taken into account in order to reproduce the correct magnetic
moment of $\mathrm{Tl_{2}Mn_{2}O_{7}}$. Similarly, these results suggest that hybridization between
Os-O(1) and Cd-O(2) may be important in understanding the magnetism in this compound.

\begin{figure}
\includegraphics[width = 3 in]{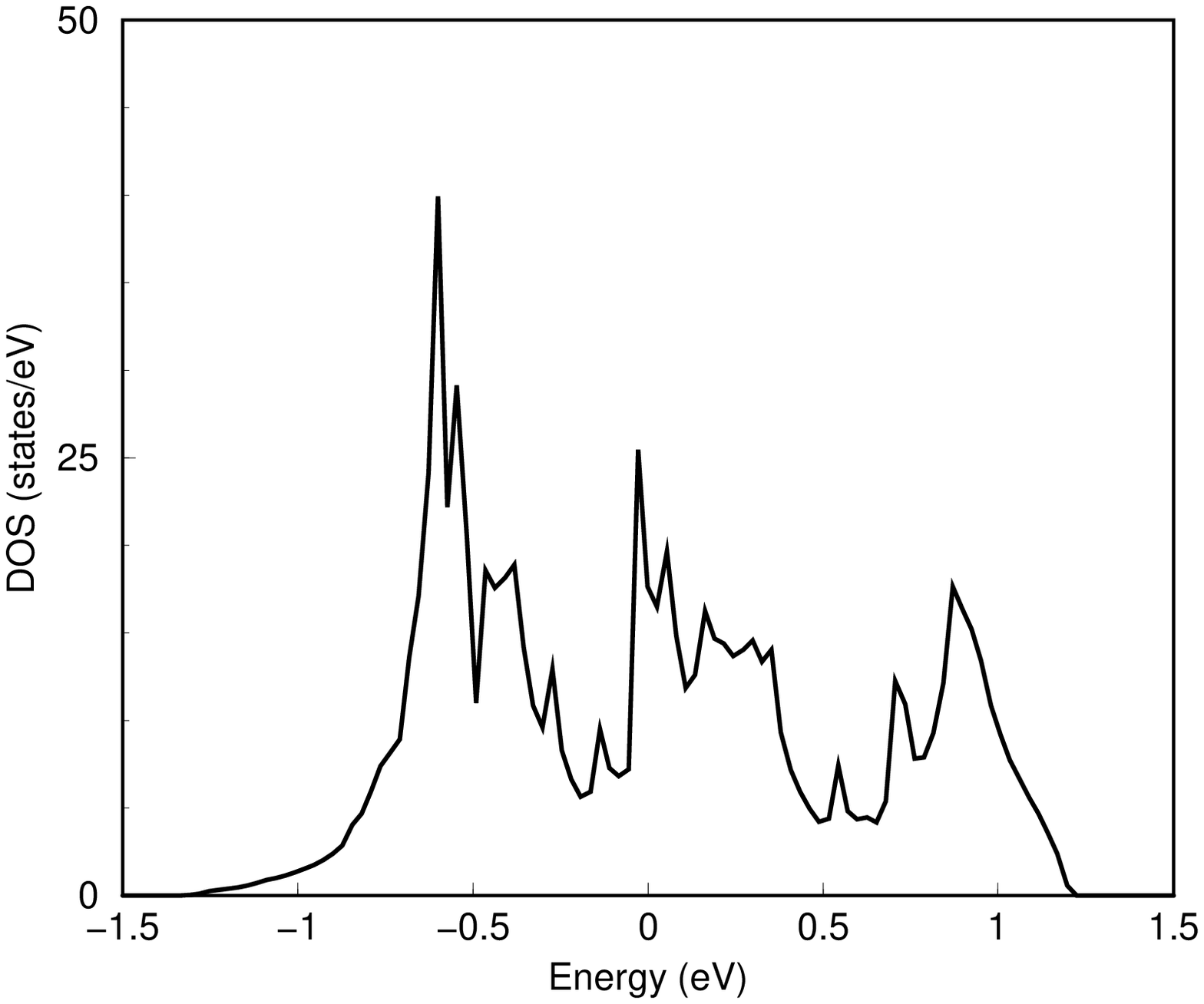}
\caption{Total DOS of $\mathrm{Cd_{2}Os_{2}O_{7}}$ near $E_F$.}
\end{figure}

\section{Conclusion}

In this work we have characterized the metal-insulator transition in $\mathrm{Cd_2Os_2O_7}$ using a
variety of experimental techniques and have argued that a coherent picture of the transition emerges
if the data are interpreted in terms of a Slater transition.  Although the possibility of a Slater
transition has occasionally been mentioned in the study of metal-insulator transitions
(\textit{e.g.}, $\mathrm{V_2O_3}$ under pressure \cite{Carter91} \cite{V2O3carter}),
$\mathrm{Cd_2Os_2O_7}$ appears to be the first well-documented example of a pure Slater transition
and therefore merits further study.  Particularly needed are theoretical studies in the intermediate
coupling regime ($U/W \approx 1$). This is the regime that characterizes many $4d/5d$ materials and
promises to yield many intriguing discoveries in the years ahead.


%
%

%

\begin{acknowledgments}
We thank J. He for assistance with the Hall measurements and S. E. Nagler, J. O. Sofo, S. A. Trugman,
G. Murthy, and J. Zaanen for helpful conversations.  Oak Ridge National laboratory is managed by
UT-Battelle, LLC, for the U.S. Department of Energy under contract DE-AC05-00OR22725.
\end{acknowledgments}

\bibliography{cdosbib.tex}

\begin{thebibliography}{10}
\expandafter\ifx\csname bibnamefont\endcsname\relax
  \def\bibnamefont#1{#1}\fi
\expandafter\ifx\csname bibfnamefont\endcsname\relax
  \def\bibfnamefont#1{#1}\fi
\expandafter\ifx\csname url\endcsname\relax
  \def\url#1{\texttt{#1}}\fi
\expandafter\ifx\csname urlprefix\endcsname\relax\def\urlprefix{URL }\fi
\providecommand{\bibinfo}[2]{#2}
\providecommand{\eprint}[2][]{\url{#2}}

\bibitem{Sleight66}
\bibinfo{author}{\bibfnamefont{A.}~\bibnamefont{Sleight}} \bibnamefont{and}
  \bibinfo{author}{\bibfnamefont{J.}~\bibnamefont{Gillson}},
  \bibinfo{journal}{Solid State Comm.} \textbf{\bibinfo{volume}{4}},
  \bibinfo{pages}{601} (\bibinfo{year}{1966}).

\bibitem{Maeno99}
\bibinfo{author}{\bibfnamefont{Y.}~\bibnamefont{Maeno}},
  \bibinfo{author}{\bibfnamefont{S.}~\bibnamefont{Nakatsuji}},
  \bibnamefont{and} \bibinfo{author}{\bibfnamefont{S.}~\bibnamefont{Ikeda}},
  \bibinfo{journal}{Mat. Science and Engineering}
  \textbf{\bibinfo{volume}{B63}}, \bibinfo{pages}{70} (\bibinfo{year}{1999}).

\bibitem{Cox83}
\bibinfo{author}{\bibfnamefont{P.}~\bibnamefont{Cox}},
  \bibinfo{author}{\bibfnamefont{R.}~\bibnamefont{Egdell}},
  \bibinfo{author}{\bibfnamefont{J.}~\bibnamefont{Goodenough}},
  \bibinfo{author}{\bibfnamefont{A.}~\bibnamefont{Hamnett}}, \bibnamefont{and}
  \bibinfo{author}{\bibfnamefont{C.}~\bibnamefont{Naish}}, \bibinfo{journal}{J.
  Phys. C} \textbf{\bibinfo{volume}{16}}, \bibinfo{pages}{6221}
  (\bibinfo{year}{1983}).

\bibitem{Cao98}
\bibinfo{author}{\bibfnamefont{G.}~\bibnamefont{Cao}},
  \bibinfo{author}{\bibfnamefont{J.}~\bibnamefont{Bolivar}},
  \bibinfo{author}{\bibfnamefont{S.}~\bibnamefont{McCall}},
  \bibinfo{author}{\bibfnamefont{J.~E.} \bibnamefont{Crow}}, \bibnamefont{and}
  \bibinfo{author}{\bibfnamefont{R.~P.} \bibnamefont{Guertin}},
  \bibinfo{journal}{Phys. Rev. B} \textbf{\bibinfo{volume}{57}},
  \bibinfo{pages}{R11039} (\bibinfo{year}{1998}).

\bibitem{Mandrus99}
\bibinfo{author}{\bibfnamefont{D.}~\bibnamefont{Mandrus}},
  \bibinfo{author}{\bibfnamefont{V.}~\bibnamefont{Keppens}}, \bibnamefont{and}
  \bibinfo{author}{\bibfnamefont{B.~C.} \bibnamefont{Chakoumakos}},
  \bibinfo{journal}{Mat. Res. Bull.} \textbf{\bibinfo{volume}{34}},
  \bibinfo{pages}{1013} (\bibinfo{year}{1999}).

\bibitem{Longo71}
\bibinfo{author}{\bibfnamefont{J.~M.} \bibnamefont{Longo}},
  \bibinfo{author}{\bibfnamefont{J.~A.} \bibnamefont{Kafalas}},
  \bibnamefont{and} \bibinfo{author}{\bibfnamefont{R.~J.}
  \bibnamefont{Arnott}}, \bibinfo{journal}{J. Solid State Chem.}
  \textbf{\bibinfo{volume}{3}}, \bibinfo{pages}{174} (\bibinfo{year}{1971}).

\bibitem{Maeno94}
\bibinfo{author}{\bibfnamefont{Y.}~\bibnamefont{Maeno}},
  \bibinfo{author}{\bibfnamefont{H.}~\bibnamefont{Hashimoto}},
  \bibinfo{author}{\bibfnamefont{K.}~\bibnamefont{Yoshida}},
  \bibinfo{author}{\bibfnamefont{S.}~\bibnamefont{Nishizaki}},
  \bibinfo{author}{\bibfnamefont{T.}~\bibnamefont{Fujita}},
  \bibinfo{author}{\bibfnamefont{J.~G.} \bibnamefont{Bednorz}},
  \bibnamefont{and}
  \bibinfo{author}{\bibfnamefont{F.}~\bibnamefont{Lichtenberg}},
  \bibinfo{journal}{Nature} \textbf{\bibinfo{volume}{372}},
  \bibinfo{pages}{532} (\bibinfo{year}{1994}).

\bibitem{Bernhard99}
\bibinfo{author}{\bibfnamefont{C.}~\bibnamefont{Bernhard}},
  \bibinfo{author}{\bibfnamefont{J.~L.} \bibnamefont{Tallon}},
  \bibinfo{author}{\bibfnamefont{C.}~\bibnamefont{Niedermayer}},
  \bibinfo{author}{\bibfnamefont{T.}~\bibnamefont{Blasius}},
  \bibinfo{author}{\bibfnamefont{A.}~\bibnamefont{Golnik}},
  \bibinfo{author}{\bibfnamefont{E.}~\bibnamefont{Br{\"u}cher}},
  \bibinfo{author}{\bibfnamefont{R.~K.} \bibnamefont{Kremer}},
  \bibinfo{author}{\bibfnamefont{D.~R.} \bibnamefont{Noakes}},
  \bibinfo{author}{\bibfnamefont{C.~E.} \bibnamefont{Stronach}},
  \bibnamefont{and} \bibinfo{author}{\bibfnamefont{E.~J.}
  \bibnamefont{Ansaldo}}, \bibinfo{journal}{Phys. Rev. B}
  \textbf{\bibinfo{volume}{59}}, \bibinfo{pages}{14099} (\bibinfo{year}{1999}).

\bibitem{Lynn2000}
\bibinfo{author}{\bibfnamefont{J.~W.} \bibnamefont{Lynn}},
  \bibinfo{author}{\bibfnamefont{B.}~\bibnamefont{Keimer}},
  \bibinfo{author}{\bibfnamefont{C.}~\bibnamefont{Ulrich}},
  \bibinfo{author}{\bibfnamefont{C.}~\bibnamefont{Bernhard}}, \bibnamefont{and}
  \bibinfo{author}{\bibfnamefont{J.~L.} \bibnamefont{Tallon}},
  \bibinfo{journal}{Phys. Rev. B} \textbf{\bibinfo{volume}{61}},
  \bibinfo{pages}{R14964} (\bibinfo{year}{2000}).

\bibitem{Sleight74}
\bibinfo{author}{\bibfnamefont{A.~W.} \bibnamefont{Sleight}},
  \bibinfo{author}{\bibfnamefont{J.~L.} \bibnamefont{Gillson}},
  \bibinfo{author}{\bibfnamefont{J.~F.} \bibnamefont{Weiher}},
  \bibnamefont{and} \bibinfo{author}{\bibfnamefont{W.}~\bibnamefont{Bindloss}},
  \bibinfo{journal}{Solid State Comm.} \textbf{\bibinfo{volume}{14}},
  \bibinfo{pages}{357} (\bibinfo{year}{1974}).

\bibitem{Slater51}
\bibinfo{author}{\bibfnamefont{J.~C.} \bibnamefont{Slater}},
  \bibinfo{journal}{Phys. Rev.} \textbf{\bibinfo{volume}{82}},
  \bibinfo{pages}{538} (\bibinfo{year}{1951}).

\bibitem{Matsubara54}
\bibinfo{author}{\bibfnamefont{T.}~\bibnamefont{Matsubara}} \bibnamefont{and}
  \bibinfo{author}{\bibfnamefont{Y.}~\bibnamefont{Yokota}}, in
  \emph{\bibinfo{booktitle}{Proc. Int. Conf. Theor. Phys., Kyoto-Tokyo 1953}}
  (\bibinfo{publisher}{Sci. Council Japan}, \bibinfo{address}{Tokyo},
  \bibinfo{year}{1954}), p. \bibinfo{pages}{693}.

\bibitem{DesCloizeaux59}
\bibinfo{author}{\bibfnamefont{J.}~\bibnamefont{{Des Cloizeaux}}},
  \bibinfo{journal}{J. Phys. Radium, Paris} \textbf{\bibinfo{volume}{20}},
  \bibinfo{pages}{606} (\bibinfo{year}{1959}).

\bibitem{Fazekas99}
\bibinfo{author}{\bibfnamefont{P.}~\bibnamefont{Fazekas}},
  \emph{\bibinfo{title}{Lecture Notes on Electron Correlation and Magnetism}}
  (\bibinfo{publisher}{World Scientific}, \bibinfo{address}{Singapore},
  \bibinfo{year}{1999}).

\bibitem{Mott90}
\bibinfo{author}{\bibfnamefont{N.~F.} \bibnamefont{Mott}},
  \emph{\bibinfo{title}{Metal-Insulator Transitions}}
  (\bibinfo{publisher}{Taylor and Francis}, \bibinfo{address}{London},
  \bibinfo{year}{1990}).

\bibitem{Imada98}
\bibinfo{author}{\bibfnamefont{M.}~\bibnamefont{Imada}},
  \bibinfo{author}{\bibfnamefont{A.}~\bibnamefont{Fujimori}}, \bibnamefont{and}
  \bibinfo{author}{\bibfnamefont{Y.}~\bibnamefont{Tokura}},
  \bibinfo{journal}{Rev. Mod. Phys.} \textbf{\bibinfo{volume}{70}},
  \bibinfo{pages}{1039} (\bibinfo{year}{1998}).

\bibitem{Subramanian83}
\bibinfo{author}{\bibfnamefont{M.~A.} \bibnamefont{Subramanian}},
  \bibinfo{author}{\bibfnamefont{G.}~\bibnamefont{Aravamudan}},
  \bibnamefont{and} \bibinfo{author}{\bibfnamefont{G.~V.~S.}
  \bibnamefont{Rao}}, \bibinfo{journal}{Prog. Solid State Chem.}
  \textbf{\bibinfo{volume}{15}}, \bibinfo{pages}{55} (\bibinfo{year}{1983}).

\bibitem{Donohue65}
\bibinfo{author}{\bibfnamefont{P.~C.} \bibnamefont{Donohue}},
  \bibinfo{author}{\bibfnamefont{J.~M.} \bibnamefont{Longo}},
  \bibinfo{author}{\bibfnamefont{R.~D.} \bibnamefont{Rosenstein}},
  \bibnamefont{and} \bibinfo{author}{\bibfnamefont{L.}~\bibnamefont{Katz}},
  \bibinfo{journal}{Inorg. Chem.} \textbf{\bibinfo{volume}{4}},
  \bibinfo{pages}{1152} (\bibinfo{year}{1965}).

\bibitem{Wang98}
\bibinfo{author}{\bibfnamefont{R.}~\bibnamefont{Wang}} \bibnamefont{and}
  \bibinfo{author}{\bibfnamefont{A.~W.} \bibnamefont{Sleight}},
  \bibinfo{journal}{Mat. Res. Bull.} \textbf{\bibinfo{volume}{33}},
  \bibinfo{pages}{1005} (\bibinfo{year}{1998}).

\bibitem{Sleight76}
\bibinfo{author}{\bibfnamefont{A.~W.} \bibnamefont{Sleight}} \bibnamefont{and}
  \bibinfo{author}{\bibfnamefont{J.~D.} \bibnamefont{Bierlein}},
  \bibinfo{journal}{Solid State Comm.} \textbf{\bibinfo{volume}{18}},
  \bibinfo{pages}{163} (\bibinfo{year}{1976}).

\bibitem{Chako84}
\bibinfo{author}{\bibfnamefont{B.~C.} \bibnamefont{Chakoumakos}},
  \bibinfo{journal}{J. Solid State Chem} \textbf{\bibinfo{volume}{53}},
  \bibinfo{pages}{120} (\bibinfo{year}{1984}).

\bibitem{Allen96}
\bibinfo{author}{\bibfnamefont{P.~B.} \bibnamefont{Allen}},
  \bibinfo{author}{\bibfnamefont{H.}~\bibnamefont{Berger}},
  \bibinfo{author}{\bibfnamefont{O.}~\bibnamefont{Chauvet}},
  \bibinfo{author}{\bibfnamefont{L.}~\bibnamefont{Forro}},
  \bibinfo{author}{\bibfnamefont{T.}~\bibnamefont{Jarlborg}},
  \bibinfo{author}{\bibfnamefont{A.}~\bibnamefont{Junod}},
  \bibinfo{author}{\bibfnamefont{B.}~\bibnamefont{Revaz}}, \bibnamefont{and}
  \bibinfo{author}{\bibfnamefont{G.}~\bibnamefont{Santi}},
  \bibinfo{journal}{Phys. Rev. B.} \textbf{\bibinfo{volume}{53}},
  \bibinfo{pages}{4393} (\bibinfo{year}{1996}).

\bibitem{Anderson56}
\bibinfo{author}{\bibfnamefont{P.~W.} \bibnamefont{Anderson}},
  \bibinfo{journal}{Phys. Rev.} \textbf{\bibinfo{volume}{102}},
  \bibinfo{pages}{1008} (\bibinfo{year}{1956}).

\bibitem{Ramirez94}
\bibinfo{author}{\bibfnamefont{A.~P.} \bibnamefont{Ramirez}},
  \bibinfo{journal}{Ann. Rev. Mater. Sci.} \textbf{\bibinfo{volume}{24}},
  \bibinfo{pages}{453} (\bibinfo{year}{1994}).

\bibitem{Pike77}
\bibinfo{author}{\bibfnamefont{G.~E.} \bibnamefont{Pike}} \bibnamefont{and}
  \bibinfo{author}{\bibfnamefont{C.~H.} \bibnamefont{Seager}},
  \bibinfo{journal}{J. Appl. Phys.} \textbf{\bibinfo{volume}{48}},
  \bibinfo{pages}{5152} (\bibinfo{year}{1977}).

\bibitem{Gruner94}
\bibinfo{author}{\bibfnamefont{G.}~\bibnamefont{Gr{\"u}ner}},
  \bibinfo{journal}{Rev. Mod. Physics} \textbf{\bibinfo{volume}{66}},
  \bibinfo{pages}{1} (\bibinfo{year}{1994}).

\bibitem{Klein96}
\bibinfo{author}{\bibfnamefont{L.}~\bibnamefont{Klein}},
  \bibinfo{author}{\bibfnamefont{J.~S.} \bibnamefont{Dodge}},
  \bibinfo{author}{\bibfnamefont{C.~H.} \bibnamefont{Ahn}},
  \bibinfo{author}{\bibfnamefont{G.~J.} \bibnamefont{Snyder}},
  \bibinfo{author}{\bibfnamefont{T.~H.} \bibnamefont{Geballe}},
  \bibinfo{author}{\bibfnamefont{M.~R.} \bibnamefont{Beasley}},
  \bibnamefont{and}
  \bibinfo{author}{\bibfnamefont{A.}~\bibnamefont{Kapitulnik}},
  \bibinfo{journal}{Phys. Rev. Lett.} \textbf{\bibinfo{volume}{77}},
  \bibinfo{pages}{2774} (\bibinfo{year}{1996}).

\bibitem{Sullow98}
\bibinfo{author}{\bibfnamefont{S.}~\bibnamefont{Sullow}},
  \bibinfo{author}{\bibfnamefont{I.}~\bibnamefont{Prasad}},
  \bibinfo{author}{\bibfnamefont{M.~C.} \bibnamefont{Aronson}},
  \bibinfo{author}{\bibfnamefont{J.~L.} \bibnamefont{Sarrao}},
  \bibinfo{author}{\bibfnamefont{Z.}~\bibnamefont{Fisk}},
  \bibinfo{author}{\bibfnamefont{D.}~\bibnamefont{Hristova}},
  \bibinfo{author}{\bibfnamefont{A.}~\bibnamefont{Lacerda}},
  \bibinfo{author}{\bibfnamefont{M.~F.} \bibnamefont{Hundley}},
  \bibinfo{author}{\bibfnamefont{A.}~\bibnamefont{Vigliante}},
  \bibnamefont{and} \bibinfo{author}{\bibfnamefont{D.}~\bibnamefont{Gibbs}},
  \bibinfo{journal}{Phys. Rev. B} \textbf{\bibinfo{volume}{77}},
  \bibinfo{pages}{5860} (\bibinfo{year}{1998}).

\bibitem{McCauley73}
\bibinfo{author}{\bibfnamefont{R.~A.} \bibnamefont{McCauley}},
  \bibinfo{journal}{J. Opt. Soc. Amer.} \textbf{\bibinfo{volume}{63}},
  \bibinfo{pages}{721} (\bibinfo{year}{1973}).

\bibitem{Blacklock79}
\bibinfo{author}{\bibfnamefont{K.}~\bibnamefont{Blacklock}} \bibnamefont{and}
  \bibinfo{author}{\bibfnamefont{H.~W.} \bibnamefont{White}},
  \bibinfo{journal}{J. Chem. Phys.} \textbf{\bibinfo{volume}{71}},
  \bibinfo{pages}{5287} (\bibinfo{year}{1979}).

\bibitem{Arrott66}
\bibinfo{author}{\bibfnamefont{A.}~\bibnamefont{Arrott}}, in
  \emph{\bibinfo{booktitle}{Magnetism}}, edited by
  \bibinfo{editor}{\bibfnamefont{G.~T.} \bibnamefont{Rado}} \bibnamefont{and}
  \bibinfo{editor}{\bibfnamefont{H.}~\bibnamefont{Suhl}}
  (\bibinfo{publisher}{Academic Press}, \bibinfo{address}{New York},
  \bibinfo{year}{1966}), vol. \bibinfo{volume}{IIB}.

\bibitem{Reimers91}
\bibinfo{author}{\bibfnamefont{J.~N.} \bibnamefont{Reimers}},
  \bibinfo{author}{\bibfnamefont{A.~J.} \bibnamefont{Berlinsky}},
  \bibnamefont{and} \bibinfo{author}{\bibfnamefont{A.-C.} \bibnamefont{Shi}},
  \bibinfo{journal}{Phys. Rev. B} \textbf{\bibinfo{volume}{43}},
  \bibinfo{pages}{865} (\bibinfo{year}{1991}).

\bibitem{sales97}
\bibinfo{author}{\bibfnamefont{B.~C.} \bibnamefont{Sales}},
  \bibinfo{author}{\bibfnamefont{D.}~\bibnamefont{Mandrus}},
  \bibinfo{author}{\bibfnamefont{B.~C.} \bibnamefont{Chakoumakos}},
  \bibinfo{author}{\bibfnamefont{V.}~\bibnamefont{Keppens}}, \bibnamefont{and}
  \bibinfo{author}{\bibfnamefont{J.~R.} \bibnamefont{Thompson}},
  \bibinfo{journal}{Phys. Rev. B} \textbf{\bibinfo{volume}{56}},
  \bibinfo{pages}{15081} (\bibinfo{year}{1997}).

\bibitem{Tsuda91}
\bibinfo{author}{\bibfnamefont{N.}~\bibnamefont{Tsuda}},
  \bibinfo{author}{\bibfnamefont{K.}~\bibnamefont{Nasu}},
  \bibinfo{author}{\bibfnamefont{A.}~\bibnamefont{Yanase}}, \bibnamefont{and}
  \bibinfo{author}{\bibfnamefont{K.}~\bibnamefont{Siratori}},
  \emph{\bibinfo{title}{Electronic Conduction in Oxides}}
  (\bibinfo{publisher}{Springer-Verlag}, \bibinfo{address}{Berlin},
  \bibinfo{year}{1991}).

\bibitem{Chaikin90}
\bibinfo{author}{\bibfnamefont{P.~M.} \bibnamefont{Chaikin}}, in
  \emph{\bibinfo{booktitle}{Organic Superconductivity}}, edited by
  \bibinfo{editor}{\bibfnamefont{V.~Z.} \bibnamefont{Krezin}} \bibnamefont{and}
  \bibinfo{editor}{\bibfnamefont{W.~A.} \bibnamefont{Little}}
  (\bibinfo{publisher}{Plenum Press}, \bibinfo{address}{New York},
  \bibinfo{year}{1990}).

\bibitem{Schultz95}
\bibinfo{author}{\bibfnamefont{W.~W.} \bibnamefont{Schultz}} \bibnamefont{and}
  \bibinfo{author}{\bibfnamefont{P.~B.} \bibnamefont{Allen}},
  \bibinfo{journal}{Phys. Rev. B} \textbf{\bibinfo{volume}{56}},
  \bibinfo{pages}{7994} (\bibinfo{year}{1995}).

\bibitem{V2O3carter}
\bibinfo{author}{\bibfnamefont{S.~A.} \bibnamefont{Carter}},
  \bibinfo{author}{\bibfnamefont{T.~F.} \bibnamefont{Rosenbaum}},
  \bibinfo{author}{\bibfnamefont{M.}~\bibnamefont{Lu}},
  \bibinfo{author}{\bibfnamefont{H.~M.} \bibnamefont{Jaeger}},
  \bibinfo{author}{\bibfnamefont{P.}~\bibnamefont{Metcalf}},
  \bibinfo{author}{\bibfnamefont{J.~M.} \bibnamefont{Honig}}, \bibnamefont{and}
  \bibinfo{author}{\bibfnamefont{J.}~\bibnamefont{Spalek}},
  \bibinfo{journal}{Phys. Rev. B} \textbf{\bibinfo{volume}{49}},
  \bibinfo{pages}{7898} (\bibinfo{year}{1994}).

\bibitem{blaha}
\bibinfo{author}{\bibfnamefont{P.}~\bibnamefont{Blaha}},
  \bibinfo{author}{\bibfnamefont{K.}~\bibnamefont{Schwarz}},
  \bibinfo{author}{\bibfnamefont{P.}~\bibnamefont{Sorantin}}, \bibnamefont{and}
  \bibinfo{author}{\bibfnamefont{S.~B.} \bibnamefont{Trickey}},
  \bibinfo{journal}{Comput. Phys. Commun.} \textbf{\bibinfo{volume}{49}},
  \bibinfo{pages}{399} (\bibinfo{year}{1990}).

\bibitem{perdew}
\bibinfo{author}{\bibfnamefont{J.~P.} \bibnamefont{Perdew}},
  \bibinfo{author}{\bibfnamefont{S.}~\bibnamefont{Burke}}, , \bibnamefont{and}
  \bibinfo{author}{\bibfnamefont{M.}~\bibnamefont{Ernzerhof}},
  \bibinfo{journal}{Phys. Rev. Lett.} \textbf{\bibinfo{volume}{77}},
  \bibinfo{pages}{3865} (\bibinfo{year}{1996}).

\bibitem{singh}
\bibinfo{author}{\bibfnamefont{D.~J.} \bibnamefont{Singh}},
  \bibinfo{journal}{Phys. Rev. B} \textbf{\bibinfo{volume}{77}},
  \bibinfo{pages}{3865} (\bibinfo{year}{1997}).

\bibitem{Carter91}
\bibinfo{author}{\bibfnamefont{S.~A.} \bibnamefont{Carter}},
  \bibinfo{author}{\bibfnamefont{J.}~\bibnamefont{Yang}},
  \bibinfo{author}{\bibfnamefont{T.~F.} \bibnamefont{Rosenbaum}},
  \bibinfo{author}{\bibfnamefont{J.}~\bibnamefont{Spalek}}, \bibnamefont{and}
  \bibinfo{author}{\bibfnamefont{J.~M.} \bibnamefont{Honig}},
  \bibinfo{journal}{Phys. Rev. B} \textbf{\bibinfo{volume}{43}},
  \bibinfo{pages}{607} (\bibinfo{year}{1991}).

\end{thebibliography}

\end{document}